\documentclass[12pt, reqno]{amsart}
\usepackage{latexsym, enumerate, imakeidx,mathrsfs, color,amscd,amsmath,amsfonts,amssymb,geometry,color,amsthm, float}
\usepackage[english]{minitoc}
\usepackage[bookmarksnumbered, colorlinks]{hyperref}
\hypersetup{colorlinks=true,linkcolor=black, anchorcolor=green, citecolor=cyan, urlcolor=red, filecolor=magenta, pdftoolbar=true}

\newtheorem{theorem}{Theorem}[section]
\newtheorem{lemma}[theorem]{Lemma}
\newtheorem{proposition}[theorem]{Proposition}
\newtheorem{corollary}[theorem]{Corollary}
\theoremstyle{definition}
\newtheorem{definition}[theorem]{Definition}
\newtheorem{example}[theorem]{Example}

\theoremstyle{remark}
\newtheorem{remark}[theorem]{Remark}
\numberwithin{equation}{section}

\DeclareMathOperator{\tr}{tr\,\!}

\newcommand{\mcl}{\mathcal}
\newcommand{\hs}{\mathcal{H}}
\newcommand{\vna}{\mathcal{N}}
\newcommand{\identity}{\mathds{1}}
\newcommand{\ent}[2]{D ( #1 || #2 )}
\newcommand{\entA}[3]{D_{#1}(#2||#3)}
\newcommand{\entdA}[3]{D_{#1}^d(#2||#3)}

\newcommand{\entbA}[3]{D_{#1}^{\bullet}(#2||#3)}
\newcommand{\entrA}[3]{D_{#1}^{\mcl{R}}(#2||#3)}
\newcommand{\entXA}[4]{D_{#2}^{#1}(#3||#4)}
\newcommand{\bs}[2]{\hat{D} ( #1 || #2 )}

\newcommand{\condexp}{\mathds{E}}
\newcommand{\hsbraket}[2]{\langle{#1,#2}\rangle}

\usepackage{physics, dsfont, tikz, enumitem}
\definecolor{figRegionA}{HTML}{C8E1F0}
\definecolor{figRegionB}{HTML}{F7DDA3}
\definecolor{figRegionC}{HTML}{CDEADF}
\definecolor{figConditioning}{HTML}{EAD9E7}

\makeatletter
\renewcommand{\subsection}{\@startsection{subsection}{2}%
  \z@{.5\linespacing\@plus.7\linespacing}{.3\linespacing}%
  {\normalfont\bfseries}}
\renewcommand{\subsubsection}{\@startsection{subsubsection}{3}%
  \z@{.5\linespacing\@plus.7\linespacing}{.3\linespacing}%
  {\normalfont\itshape}}
\makeatother

\usepackage[url=false]{biblatex}
\begin{document}

\title{Quantum conditional relative entropy and quasi-factorization}

\author[A. Lucia]{Angelo Lucia}

\address{
Dipartimento di Matematica\\ Politecnico di Milano\\
Piazza Leonardo da Vinci 32 \\ 20133 Milano\\
Italy}
\email{angelo.lucia@polimi.it}

\author[A. Capel]{Ángela Capel}
\address{
Department of Applied Mathematics and Theoretical Physics, University of Cambridge,
Wilberforce Road, Cambridge CB3 0WA, United Kingdom, and Fachbereich Mathematik, Universität Tübingen, 72076 Tübingen, Germany}
\email{ac2722@cam.ac.uk}





\begin{abstract}
One of the characteristic features that distinguishes quantum systems from classical ones is the inability, except for some special cases, of defining a conditional expectation map for a given quantum state, namely a map that restricts observables to a given sub-region (or more generally, a sub-algebra of observables) while preserving expectation values.
Of the many consequences of this lack of conditional expectation, we consider the problem of defining the quantum conditional relative entropy, i.e., the relative entropy between two quantum states conditioned to a sub-region or sub-algebra. We discuss how, when the conditional expectation map is not defined, there are multiple, non-equivalent ways to define a quantum conditional relative entropy quantity, and we present some of their properties.

Classical conditional relative entropy has played a fundamental role in many applications to spin systems, and in particular in proving logarithmic Sobolev inequalities. A key feature of conditional relative entropy that intervenes in these results is its quasi-factorization or approximate tensorization property: informally, the relative entropy is almost super-additive over weakly-dependent sub-algebras.
It is then not surprising that the key to obtaining similar applications in the quantum setting, such as quantum logarithmic Sobolev inequalities in spin systems, has been to obtain approximate factorization results for the quantum conditional relative entropy,
with the added complication that there is not a unique definition of conditional relative entropy to work with. We review the recent results regarding quasi-factorization of the quantum conditional relative entropy, in strong and weak forms.
\end{abstract}

\maketitle

\section{Introduction}

Classical statistical mechanics uses the tools of probability theory to describe the macroscopic behavior of collections of many elementary particles. On a very high level, it describes the physical system through  a probability space $(\Omega, \mcl{F}, \mu)$, where the sample space $\Omega$ represents the phase space of the system, and the $\sigma$-algebra $\mcl{F}$ represents possible events, which can happen with a certain probability described by the distribution $\mu$.
Observables of the system are given by elements of $L_0(\mcl{F})$, the space of $\mcl{F}$-measurable real functions, and for an integrable observable $f\in L_1(\mcl{F},\mu)$, its expected value is given by $\mu(f)$.
If $\nu$ is another probability measure on $\mcl{F}$, which is absolutely continuous with respect to $\mu$, then it is possible to define the \emph{relative entropy} of $\nu$ with respect to $\mu$, also known as the Kullback–Leibler divergence between $\nu$ and $\mu$, as
\begin{equation}
    H(\nu||\mu) = \nu\qty( \log \dv{\nu}{\mu} ) = \int \dd\nu \log \dv{\nu}{\mu}.
\end{equation}
Note that if both $\nu$ and $\mu$ are absolutely continuous with respect to some reference probability measure $\lambda$, with $\nu=p \lambda$ and $\mu=q\lambda$ for some $p,q \in L_1(\lambda)$, then the relative entropy takes the familiar form
\begin{equation}
H(\nu||\mu) = H(p || q) = \int\dd\lambda\, p\, (\log p - \log q).
\end{equation}

If we consider instead quantum systems, this picture carries over with some natural adjustments. Given a finite dimensional\footnote{We will not consider infinite dimensional settings in this chapter, but the theory can be extended to that case, with some work.} Hilbert space $\hs$, the space of \emph{observables} of the system is given by (self-adjoint elements of) $\mcl{B}(\hs)$, and their expected value is fixed by the choice of a positive semi-definite operator $\sigma\in \mcl{B}(\hs)$ with unit trace, via Born's rule: $\expval{X}_\sigma = \Tr \sigma X$. If we have another such operator $\rho$, which is absolutely continuous with respect to $\sigma$ in the sense that $\expval{X}_\sigma = 0 \Rightarrow \expval{X}_\rho = 0$ for every positive $X$, one can define \emph{Umegaki's relative entropy} as
\begin{equation}
    \ent \rho \sigma = \expval{\log \rho - \log \sigma}_\rho = \Tr \rho(\log \rho - \log \sigma).
\end{equation}

The parallelism between classical and quantum probability breaks down if we want to \emph{condition} the state of the system with respect to some sub-algebra. Classically, this poses no problem: if $\mcl{G}$ is a sub-$\sigma$-algebra of $\mcl{F}$, then for each probability measure $\mu$ there exists a unique (up to $\mu$-almost-everywhere equality) \emph{conditional expectation} $\condexp_\mu[\cdot | \mcl{G}] : L_1(\mcl{F},\mu) \to L_1(\mcl{G},\mu|_\mcl{G})$ which preserves expectation values with respect to $\mu$, in the sense that
\begin{equation}\label{eq:classical-cond-exp}
     \int\dd \mu\, f g  = \int \dd \mu|_{\mcl{G}} \,  \condexp_\mu[f | \mcl{G}] \, g , \quad \forall f \in L_1(\mcl{F},\mu),\  \forall g\in L_\infty(\mcl{G},\mu|_\mcl{G}).
\end{equation}
Thanks to the existence of a conditional expectation, assuming $H(\nu \| \mu)< \infty$, it is possible to define the \emph{conditional relative entropy} between $\mu$ and $\nu$, conditioned on $\mcl{G}$:
\begin{equation}
    H_{\mcl{F}|\mcl{G}} (\nu||\mu) = \nu\qty( \log \dv{\nu}{\mu} - \log \condexp_{\mu}\qty[\dv{\nu}{\mu} \Big{|} \mcl{G}]).
\end{equation}
If again we consider the case $\nu=p \lambda$ and $\mu=q\lambda$ for some $p,q \in L_1(\lambda)$, then the conditional relative entropy can be expressed in terms of the relative entropy as follows
\begin{equation}
 H_{\mcl{F}|\mcl{G}} (\nu||\mu) = H(p || q) - H( \condexp_\lambda^*(p)|| \condexp_\lambda^*(q)) = H( p || \condexp_\mu^*(p)),
\end{equation}
where $\condexp_{\lambda}^*(p)$ (resp. $\condexp_\mu^*(p)$) denotes the density with respect to $\lambda$ of the measure $(p\lambda) \circ \condexp_\lambda$ (resp. $(p\lambda)\circ \condexp_\mu$).

In quantum systems, on the contrary, a quantum conditional expectation does not always exist (at least not with the same set of properties of the classical one). Hence, the problem of defining conditioning for quantum systems has received a lot of attention: we will not attempt to review the long history of results on this topic, but we will limit ourselves to recalling the part of the theory that we will need (in Section~\ref{sec:preliminaries}).

Instead, we will consider the following question: in the cases where a quantum conditional expectation is \emph{not} available, how can one define a \emph{quantum conditional relative entropy}? It turns out, as it is often the case for quantum generalizations of classical notions, that the answer is not unique. In this chapter, we will consider Umegaki's version of quantum relative entropy \cite{umegaki_conditional_1962}, and we will review two alternative constructions of quantum conditional relative entropy that we originally introduced in \cite{CLP18b}, bringing them into the context of the more recent developments, and discuss the similarities and differences between the two definitions.

The second half of this chapter will be dedicated to discuss a series of results regarding the so-called \emph{quasi-factorization} of the conditional relative entropy, also known as \emph{approximate tensorization} \cite{BCR22}. Exact factorization happens when we consider two independent sub-algebras, e.g., in a spin system on a lattice, operators which are supported on disjoint regions for which the  state $\sigma$ factorizes as a tensor product. In this case, the relative entropy is super-additive, in the sense that the relative entropy of the union of the regions is bounded by the sum of the conditional relative entropies for each region.
Quasi-factorization results provide extensions of these ideas to the cases where the subalgebras considered are not independent. This can happen either because the state $\sigma$ is not product, or (in the case of spin systems) because the regions considered are overlapping. In certain cases, it is still possible to recover approximate factorization results, with constants controlling ``how far'' the setting is from a purely independent case.

In classical spin systems, quasi-factorization results for the relative and conditional relative entropy have appeared numerous times, providing fundamental technical tools to prove some major results. With no pretense of being exhaustive nor complete, we list a few examples which we believe exemplify the relevance that the concept has played in the classical setting:
\begin{itemize}
    \item The work of Cesi \cite{Cesi01}, proving logarithmic Sobolev inequalities from the Dobru\-shin--\-Shlos\-man complete analyticity condition, and that of Dai Pra, Paganoni and Posta \cite{DaiPraPaganoniPosta02}, proving exponential entropy decay for unbounded spin systems where the log-Sobolev inequality fails;
    \item The works of  Marton \cite{Marton13} and Otto and Reznikoff \cite{OttoReznikoff07}, obtaining log-Sobolev inequalities from conditional relative entropy inequalities for Euclidean Gibbs measures;
    \item The works of Caputo, Menz and Tetali \cite{CaputoMenzTetali15} on approximate tensorization of entropy, and of Caputo and Parisi \cite{CaputoParisi21}, on block factorization of relative entropy, again with connections to log-Sobolev inequalities.
\end{itemize}
We also refer the reader to \cite{BlancaCaputoChenParisiStefankovicVigoda22} for a review of the connections between these results and entropy factorization.

In the quantum setting, notable major results based on the quasi-factorization of the (conditional) relative entropy have also been obtained in recent years. Among them, and again without pretense of being exhaustive, we recall the proofs of log-Sobolev inequalities in 1D spin chains \cite{BCLPR21, BCGLPR23, BCGLPR24, GaoGuoWang26, capel_modified_2026} and for quantum error correcting codes \cite{stengele_modified_2026}, entropy convergence beyond the log-Sobolev regime \cite{laracuente_self-restricting_2026}, strong data processing inequalities \cite{hirche_conditional_2026},  parallel-repetition bounds for entangled games~\cite{jain_parallel_2026}, entropic uncertainty relations and generalized strong sub-additivity conditions \cite{LaRacuente22, GaoJungeLaRacuente23}.

The chapter is organized as follows. In Section~\ref{sec:preliminaries}, we will recall some fundamental results on conditional expectations in quantum systems. In Section~\ref{sec:quantum-conditional-relative-entropy}, we will present the two definitions of quantum conditional relative entropy, and present some of their elementary properties. Section~\ref{sec:quasi-factorization} and Section~\ref{sec:weak-quasi-factorization} will be devoted to a review of recent strong and weak quasi-factorization results, respectively, with an eye towards the previously mentioned applications in quantum spin systems.

\section{Quantum conditional expectations}
\label{sec:preliminaries}

\subsection{Notation}
$\hs$ will denote a finite-dimensional Hilbert space. The set of full-rank density matrices on $\hs$ will be denoted by $\mcl{S}(\hs)$.
We equip $\mcl{B}(\hs)$ with the Hilbert-Schmidt inner product $\hsbraket{a}{b}_{2} = \Tr a^* b$, and more generally, with the $p$-th Schatten norm $\norm{a}_p = (\Tr|a|^p)^{\frac{1}{p}}$ for $p\in [1,\infty]$, with $\norm{a}_\infty$ denoting the operator norm. For a given full-rank state $\sigma \in \mcl{S}(\hs)$, we define the Gelfand–Naimark–Segal ($\sigma$-GNS) inner product as $\hsbraket{a}{b}_{GNS,\sigma} = \Tr \sigma a^* b$ and the Kubo-Martin-Schwinger ($\sigma$-KMS) inner product as $\hsbraket{a}{b}_{KMS,\sigma} = \Tr \sigma^{1/2} a^* \sigma^{1/2} b$.

\subsection{Subalgebras and conditional expectations}

Given a unital sub $*$-algebra $\vna$ of $\mcl{B}(\mcl{H})$, a \emph{Umegaki conditional expectation} \cite{Umegaki1954, Umegaki1956, Umegaki1959} is a map $\condexp: \mcl{B}(\hs) \to \mcl{N}$ which is positive, idempotent (i.e., $\condexp|_{\vna} = \operatorname{id}_{\vna}$), and satisfies the bi-module property\footnote{Note that in other works, for example in \cite{kastoryano_quantum_2016}, the term \emph{conditional expectation} denotes a map with weaker properties, in particular not necessarily satisfying the bi-module property.}:
\begin{equation}
    \condexp(abc) = a \condexp(b) c, \quad \forall a,c  \in \vna, \  \forall b \in \mcl{B}(\hs).
\end{equation}
Every such map is then necessarily unital and completely positive \cite{Tomiyama1957}.

Now consider a full-rank state $\sigma \in \mcl{S}(\hs)$.
If $\condexp$ preserves expectations with respect to $\sigma$, in the sense that $\Tr \sigma \condexp(a) = \Tr \sigma a$ for every $a\in \mcl{B}(\hs)$, then it must be the $\sigma$-GNS-orthogonal projection on $\vna$.
The latter always exists, and it is unique, due to Hilbert's projection theorem, but it might not necessarily be a conditional expectation: it will be if and only if $\sigma a \sigma^{-1} \in \vna$ for all $a\in \vna$, i.e., if $\vna$ is invariant under the modular group of $\sigma$ \cite{Takesaki1972}. Thus, for a fixed $*$-subalgebra $\vna$, we define
\begin{equation}
    \mcl{S}(\vna) = \{ \sigma \in \mcl{S}(\hs) \mid \sigma a \sigma^{-1} \in \vna \text{ for all } a\in \vna\},
\end{equation}
i.e., $\mcl{S}(\vna)$ is the set of states for which a $\sigma$-preserving conditional expectation on $\vna$ exists, and in this case we will denote it by $\condexp_\sigma$. Note that $\mcl{S}(\vna)$ is not empty, as it contains at least the maximally mixed state $\tau = \identity/\dim \hs$. The corresponding conditional expectation $\condexp_\tau$ is called the \emph{tracial} conditional expectation for $\vna$, and it is self-adjoint with respect to the Hilbert-Schmidt scalar product.
For $\sigma\in \mcl{S}(\hs)$, we denote by $\sigma_{\vna}$ the density of the restriction of $\sigma$ to $\vna$, i.e., $\sigma_{\vna} = \condexp_{\tau}(\sigma)$.

We also note that the following holds: if $\sigma \in \mcl{S}(\vna)$, then
\begin{equation}\label{eq:cond-exp-defn}
    \hsbraket{a}{b}_{GNS,\sigma} = \hsbraket{\condexp_\sigma(a)}{b}_{GNS,\sigma_{\vna}}, \quad \forall a \in \mcl{B}(\hs),\ \forall b \in \vna,
\end{equation}
which parallels \eqref{eq:classical-cond-exp} for the classical conditional expectation.

We conclude with an explicit description of $\condexp_\sigma^*$, the Hilbert-Schmidt adjoint of $\condexp_\sigma$, in the cases that it exists.
\begin{lemma}\label{lemma:sigma-N}
Suppose that $\sigma \in \mcl{S}(\vna)$. Then:
\begin{enumerate}
    \item $\sigma$ and $\sigma_{\vna}$ commute;
    \item $\sigma \sigma_{\vna}^{-1}$ commutes with every element of $\vna$;
    \item $\condexp^*_\sigma(\rho) = \sigma \sigma_{\vna}^{-1} \condexp_\tau(\rho) = \sigma \sigma_{\vna}^{-1} \rho_\vna$.
\end{enumerate}
\end{lemma}
\begin{proof}
Since $\vna$ is a finite dimensional von Neumann algebra, up to a unitary change of basis it is given by a direct sum of full matrix algebras, i.e., there exists a decomposition of $\hs$ and $\vna$ of the form:
\[
\hs = \bigoplus_{i} H_i \otimes K_i , \qquad
\vna = \bigoplus_i \identity_{H_i} \otimes \mcl{B}(K_i).
\]
With respect to this decomposition, we get that $\sigma \in \mcl{S}(\vna)$ if and only if
\[
\sigma = \bigoplus_i p_i \sigma^H_i \otimes \sigma^K_i ,\quad
\sigma^H_i \in \mcl{S}(H_i), \ \sigma^K_i \in \mcl{S}(K_i),\quad
p_i > 0, \sum_{i} p_i =1.
\]
Moreover, for an arbitrary state $\rho$, $\rho_{\vna}$  and $\condexp^*_\sigma(\rho)$ are given by
\[
\rho_\vna = \bigoplus_i p'_i \tau^H_i \otimes \rho^K_i, \quad
\condexp^*_\sigma(\rho) = \bigoplus_i p'_i \sigma_i^H \otimes \rho_i^K, \quad
\tau^H_i = \identity_{H_i}/\dim H_i,
\]
 where  $P_{H_i \otimes K_i}$ is the projection on the $i$-th block of the decomposition, and
\[ \rho_i^K = \frac{\Tr_{H_i}(P_{H_i \otimes K_i} \rho P_{H_i \otimes K_i})}{\Tr (P_{H_i \otimes K_i} \rho P_{H_i \otimes K_i})}, \quad p'_i = \Tr (P_{H_i \otimes K_i} \rho P_{H_i \otimes K_i}).\]

Hence, if $\sigma \in \mcl{S}(\vna)$, we have that
\begin{enumerate}
    \item $\sigma$ and $\sigma_{\vna}$ commute;
    \item $\sigma \sigma_{\vna}^{-1}$ commutes with every element of $\vna$, since
\[
\sigma\sigma_{\vna}^{-1} = \bigoplus_i (\dim H_i)\, \sigma^H_i \otimes \identity_{K_i} \in \mcl{N}'.
\]
    \item $\condexp^*_\sigma(\rho) = \oplus_i p'_i \sigma_i^H \otimes \rho_i^K = \oplus_i (\dim H_i)\sigma_i^H p'_i \tau_i^H \otimes \rho_i^K = \sigma \sigma_{\vna}^{-1} \condexp_\tau^*(\rho) = \sigma \sigma_{\vna}^{-1} \condexp_\tau(\rho)$.
\end{enumerate}
\end{proof}

\begin{remark}[Bipartite systems and product states]
Let us discuss how the above constructions specialize to the particular case when $\hs$ is a bipartite Hilbert space $\hs=\hs_{A} \otimes \hs_{B}$, and $\vna=\identity_A \otimes \mcl{B}(\hs_B)$. In this setting,
the tracial conditional expectation on $\vna$ is simply the normalized partial trace $\frac{1}{\dim \hs_A}\Tr_A$,
so that $\sigma_{\vna} = \condexp_\tau(\sigma) = (\frac{\identity_A}{\dim \hs_A})\otimes  \Tr_A \sigma = \tau_A \otimes \sigma_B$.
We also have that
\[\mcl{S}(\vna) = \{ \sigma_A \otimes \sigma_B \mid \sigma_A \in \mcl{S}(\hs_A), \sigma_B\in \mcl{S}(\hs_B)\}.\]
When $\sigma=\sigma_A \otimes \sigma_B$, the unique $\sigma$-invariant conditional expectation on $\vna$ is given by  $\condexp_\sigma(a) = \identity_A\otimes\Tr_A((\sigma_A\otimes\identity_B) a)$, and $\condexp^*_\sigma(\rho) = \sigma_A \otimes \rho_B$.
\end{remark}

\subsection{Coarse graining and recovery maps}
Since a $\sigma$-GNS-orthogonal conditional expectation on $\vna$ does not always exist, we will consider extensions of the mapping $\sigma \mapsto \condexp_\sigma$ which are defined on the whole $\mcl{S}(\hs)$, as formalized in the following definition.

\begin{definition}\label{def:coarse-graining-map}
    A function $\mcl{A}:\mcl{S}(\hs) \times \mcl{B}(\hs) \to \vna$ is said to be a \emph{coarse graining map} for $\vna$ if the following holds:
    \begin{enumerate}
        \item $\mcl{A}_\sigma: a \mapsto \mcl{A}(\sigma,a)$ is a unital, completely positive map  for every $\sigma \in \mcl{S}(\hs)$;
        \item $\Tr (\sigma \mcl{A}_\sigma(a)) = \Tr (\sigma a)$ for every $\sigma \in \mcl{S}(\hs)$;
        \item if $\sigma \in \mcl{S}(\vna)$, then $\mcl{A}_\sigma = \condexp_\sigma$, where $\condexp_\sigma$ is  the (unique) $\sigma$-invariant conditional expectation on $\vna$.
    \end{enumerate}
\end{definition}

It will be more convenient to work with a dual picture, in which we act on density matrices instead of observables, so we also define the following.

\begin{definition}\label{def:recovery-map}
    A function $\mcl{R}:\mcl{S}(\hs) \times \mcl{S}(\hs) \to \mcl{S}(\hs)$ is said to be a \emph{recovery map} for $\vna$ if the following holds:
    \begin{enumerate}
        \item\label{ax:recovery-cptp} $\mcl{R}_\sigma: \rho \mapsto \mcl{R}(\sigma,\rho)$ is trace-preserving and completely positive for every $\sigma\in \mcl{S}(\hs)$;
        \item\label{ax:recovery-reference} $\mcl{R}_\sigma(\sigma_\vna) = \sigma$ for every $\sigma \in \mcl{S}(\hs)$;
        \item\label{ax:recovery-restriction} $\mcl{R}_\sigma = \mcl{R}_\sigma \circ \condexp_\tau$ (i.e., $\mcl{R}_\sigma$ only depends on $\vna$);
    \end{enumerate}
    Moreover, we say that $\mcl{R}$ is an \emph{admissible recovery map} if it satisfies the following:
    \begin{enumerate}[resume]
        \item\label{ax:recovery-admissibility} if $\sigma \in \mcl{S}(\vna)$, then $\mcl{R}_\sigma = \condexp^*_\sigma$, where $\condexp^*_\sigma$ is the Hilbert-Schmidt adjoint of the (unique) $\sigma$-invariant conditional expectation on $\vna$.
    \end{enumerate}

\end{definition}
Admissible recovery maps are exactly the Hilbert-Schmidt dual of coarse graining maps, i.e. $\mcl{R}_\sigma = \mcl{A}_\sigma^*$.
Non-admissible recovery maps lack property~(\ref{ax:recovery-admissibility}), but they have found a use in the setting of recoverability theory (see Subsection~\ref{sec:relative-bounds}). In the following we will only work with admissible recovery maps, unless specifically said so.

An example of a coarse graining map is given by the so-called Accardi-Cecchini coarse graining map\footnote{Also known in the literature as \emph{minimal conditional expectation} (although it is not a conditional expectation in the sense of Umegaki), or \emph{block spin flip map}.} \cite{Accardi1982}: for any $\sigma \in \mcl{S}(\hs)$, there exists a unique linear map $\mcl{A}_{\sigma}^{AC}:\mcl{B}(\hs) \to \vna$ such that
\begin{equation}
    \hsbraket{a}{b}_{KMS,\sigma} = \hsbraket{\mcl{A}_{\sigma}^{AC}(a)}{b}_{KMS,\sigma_{\vna}}, \quad \forall a \in \mcl{B}(\hs),\ \forall b \in \vna.
\end{equation}
Compare this to \eqref{eq:cond-exp-defn} for the $\sigma$-GNS scalar product, which only holds for $\sigma\in \mcl{S}(\vna)$. The map $\mcl{A}_{\sigma}^{AC}$ can be written explicitly as
\begin{equation}
    \mcl{A}_{\sigma}^{AC}(a) = \sigma_{\vna}^{-1/2} \condexp_\tau( \sigma^{1/2} a \sigma^{1/2}) \sigma_{\vna}^{-1/2}.
\end{equation}
Its Hilbert-Schmidt dual is known in the literature as the \emph{Petz recovery map} $\mcl{R}_{\sigma}^{P}$ \cite{Petz1988}, explicitly given by
\begin{equation}
    \mcl{R}_{\sigma}^{P}(a) = \sigma^{1/2}\sigma_{\vna}^{-1/2} \condexp_\tau(a) \sigma_{\vna}^{-1/2}  \sigma^{1/2}.
\end{equation}
Note that in general neither $\mcl{A}_\sigma^{AC}$ nor $\mcl{R}_{\sigma}^{P}$ will be idempotent.

A family of variations on the Petz recovery map, named the \emph{rotated Petz recovery maps} \cite{junge_universal_2018, Wilde2015}, has also been introduced. They are indexed by a real parameter $t\in \mathds{R}$ and are defined as follows:
\begin{equation}
    \mcl{R}^{P,t}_\sigma(a) = \sigma^{\frac{1+it}{2}}\sigma_{\vna}^{\frac{-1-it}{2}} \condexp_\tau(a) \sigma_{\vna}^{\frac{-1+it}{2}}  \sigma^{\frac{1-it}{2}}, \quad t \in \mathds{R}.
\end{equation}
It is also possible to consider averages of rotated Petz recovery maps, a notable example being given by the following
\begin{equation}
\mcl{R}^{PT}_{\sigma}= \int_\mathds{R} \beta(t) \dd t\, \mcl{R}^{P,t}_\sigma, \qquad \beta(t) = \frac{\pi}{2(1+\cosh(\pi t))}.
\end{equation}

A result of \cite{Accardi1982} (see also \cite{petz_sufficient_1986, Petz1988}) shows that the Accardi-Cecchini map defines a coarse graining map $\mcl{A}(\sigma,a) = \mcl{A}_\sigma^{AC}(a)$ in the sense of Definition~\ref{def:coarse-graining-map}, and by duality the family of Petz maps defines an admissible recovery map $\mcl{R}(\sigma,\rho) = \mcl{R}_{\sigma}^{P}(\rho)$ in the sense of Definition~\ref{def:recovery-map}. Conditions~(\ref{ax:recovery-cptp}), (\ref{ax:recovery-reference}) and~(\ref{ax:recovery-restriction}) follow immediately from the definition, so the only non trivial condition to check is~(\ref{ax:recovery-admissibility}). For the standard Petz map, as well as for any of the rotated Petz maps $\mcl{R}^{P,t}_\sigma$ and their average $\mcl{R}^{PT}_\sigma$, it follows quite straightforwardly from Lemma~\ref{lemma:sigma-N}.

\subsection{Fixed point algebras}
\label{sec:cond-exp-coarse-graining}
For a subalgebra $\vna$ and a faithful state $\sigma \in \mcl{S}(\hs)$, a coarse graining map need not itself be a conditional expectation, since it will not necessarily be a $\sigma$-GNS orthogonal projection on $\vna$.
Nonetheless, it is possible to consider the sub-algebra of $\vna$ given by the fixed-point algebra
\[
 \mcl{F}_{\mcl{A}_\sigma}
 =\{a\in\vna:\mcl{A}_\sigma(a)=a\} \subset \vna.
\]
(This is an algebra since $\mcl{A}_\sigma$ is a unital, completely positive map with a full-rank invariant state).
A conditional expectation on $\mcl{F}_{\mcl{A}_\sigma}$ can always be extracted from $\mcl{A}_\sigma$ by defining
\[
 \condexp_\sigma^{\mcl{F}_{\mcl{A}_\sigma}}
 =\lim_{n\to\infty} \frac{1}{n} \sum_{k=1}^n(\mcl{A}_\sigma)^k,
 \qquad
 \bigl(\condexp_\sigma^{\mcl{F}_{\mcl{A}_\sigma}}\bigr)^*
 =\lim_{n\to\infty} \frac{1}{n} \sum_{k=1}^n (\mcl{R}_\sigma)^k.
\]
When $\mcl{A}_\sigma$ has a trivial peripheral spectrum, in the sense that $1$ is the only eigenvalue of modulus $1$, then one can avoid the Cesàro mean and simply consider the limit of the iterates of $\mcl{A}_\sigma$. This is the case for the Accardi-Cecchini map and its dual, the Petz recovery map.

From its definition, it follows that $\condexp_\sigma^{\mcl{F}_{\mcl{A}_\sigma}}$ is a unital, completely positive map with range equal to $\mcl{F}_{\mcl{A}_\sigma}$, which preserves expectation values with respect to $\sigma$. Moreover, since the fixed point algebra lives in the multiplicative domain of $\mcl{A}_\sigma$ \cite{choi_schwarz_1974}, we have that
\[
\mcl{A}^k_\sigma(abc) = a \mcl{A}^k_\sigma(b) c, \quad \forall a,c \in \mcl{F}_{\mcl{A}_\sigma}, \quad \forall b \in \mcl{B}(\hs), \quad \forall k \ge 1,
\]
hence $\condexp_\sigma^{\mcl{F}_{\mcl{A}_\sigma}}$ satisfies the bi-module property and it is a $\sigma$-invariant conditional expectation. In other words, $\sigma \in \mcl{S}(\mcl{F}_{\mcl{A}_\sigma})$.

\begin{definition}
    Let $\vna_\sigma \subset \mcl{N}$ be the largest $*$-subalgebra such that $\sigma \in \mcl{S}(\vna_\sigma)$.
\end{definition}

The previous discussion shows that $\mcl{F}_{\mcl{A}_\sigma}\subset \vna_\sigma$ for every coarse-graining map. The case of the Accardi-Cecchini map $\mcl{A}_\sigma^{AC}$ is special, since in that case we have equality:
\begin{lemma}[{\cite[Theorem 3.3]{carlen_recovery_2020}}]
    \label{lemma:accardi-cecchini-maximal}
    For the Accardi-Cecchini coarse graining map $\mcl{A}_\sigma^{AC}$, the fixed-point algebra $\mcl{F}_{\mcl{A}^{AC}_\sigma}$ is exactly equal to $\vna_\sigma$.
\end{lemma}

\section{Quantum conditional relative entropies}
\label{sec:quantum-conditional-relative-entropy}
\subsection{Two definitions of quantum conditional relative entropy}

On the set $\mcl{S}(\vna)$, we have a natural way to define what ``conditioning'' means, since we have access to the conditional expectation map $\condexp_\sigma$. This leads us to the following definition.

\begin{definition}
We define the \emph{conditional relative entropy} as
\begin{equation}
D_{\mcl{B}(\hs)|\vna} :\mcl{S}(\hs) \times  \mcl{S}(\vna) \to \mathds{R}, \qquad
        \entA{\mcl{B}(\hs)|\vna} {\rho} {\sigma} := \ent {\rho} {\condexp^*_\sigma(\rho)}.
\end{equation}
\end{definition}
Note that, as discussed in Section~\ref{sec:preliminaries}, for a given $\sigma \in \mcl{S}(\vna)$, $\condexp_\sigma$ is unique.
$D_{\mcl{B}(\hs)|\vna}$ can be re-written in a different way, whose importance will be clear soon.
\begin{lemma}[{\cite[Theorem 1.13]{ohya_quantum_2004}}]
If $\sigma \in \mcl{S}(\vna)$, then
\begin{equation}\label{eq:cond-entropy-expectations}
\entA {\mcl{B}(\hs)|\vna} {\rho} {\sigma} = \ent {\rho} {\condexp^*_\sigma(\rho)} = \ent {\rho} {\sigma} - \ent {\rho_{\vna}} {\sigma_{\vna}}
\end{equation}
\end{lemma}
This could also be seen as a corollary of Lemma~\ref{lemma:sigma-N}, since from it it follows that
$\log \condexp_\sigma^*(\rho) = \log \sigma - \log \sigma_\vna + \log \rho_{\vna}$.

If we want to extend the domain of the definition of $D_{\mcl{B}(\hs)|\vna}$ to the whole $\mcl{S}(\hs)\times \mcl{S}(\hs)$, we have to face the issue that the conditional expectation map is not available anymore when $\sigma\in \mcl{S}(\hs)\setminus \mcl{S}(\vna)$. It turns out that there are multiple, non-equivalent ways to extend the domain of the conditional relative entropy. Here we present two alternative approaches: the first is based on preserving the picture of $D_{\mcl{B}(\hs)|\vna}$ as the difference of relative entropies of the marginals, as in the r.h.s.\ of \eqref{eq:cond-entropy-expectations}; the second is instead based on preserving the picture of $D_{\mcl{B}(\hs)|\vna} = \ent{\rho}{\mcl{R}_\sigma(\rho)}$, for some admissible recovery map $\mcl{R}_\sigma$.

\begin{definition}[Conditional relative entropy by marginal difference]
We define the conditional relative entropy by marginal difference as
\begin{equation}
        \entdA {\mcl{B}(\hs)|\vna} {\rho} {\sigma} := \ent {\rho} {\sigma} - \ent {\rho_{\vna}} {\sigma_{\vna}}.
\end{equation}
\end{definition}
\begin{definition}[Conditional relative entropy by recovery]
Let $\mcl{R} : \mcl{S}(\hs)\times \mcl{S}(\hs) \to \mcl{S}(\hs)$ be an admissible recovery map for $\vna$. We then define the conditional relative entropy by recovery as
\begin{equation}
        \entrA {\mcl{B}(\hs)|\vna} {\rho} {\sigma} := \ent {\rho} {\mcl{R}_{\sigma}(\rho)}.
\end{equation}
\end{definition}
The definition of $D^{\mcl{R}}_{{\mcl{B}(\hs)|\vna}}$ was presented in \cite{CLP18b} for the specific case when $\mcl{R}_\sigma$ is the Petz recovery map, under the name of ``conditional relative entropy by expectations'' (which followed the convention of not requiring the bi-module property in the definition of conditional expectation). Instead, here we prefer to reserve this term only for its restriction to $\mcl{S}(\hs)\times \mcl{S}(\vna)$.

\subsection{The bipartite case}
In the special case where $\hs=\hs_A\otimes \hs_B$ and $\vna = \identity_A \otimes \mcl{B}(\hs_B)$, we denote $D^{\bullet}_{\mcl{B}(\hs)|\vna}$ as $D^{\bullet}_A$, where $\bullet$ is either $d$ or a choice of an admissible recovery map $\mcl{R}$.
Both definitions share a common set of properties, with $D^d_A$ satisfying a stronger property, and being characterized by it. We summarize them in the following proposition (note that in \cite{CLP18b} only the Petz recovery map family is considered, but the proof carries over to the general case).

\begin{proposition}[{\cite[Theorem 2, Proposition 6]{CLP18b}}]
Let $\entbA{A}{\cdot}{\cdot}$ denote one of $\entdA{A}{\cdot}{\cdot}$, $\entrA{A}{\cdot}{\cdot}$. They satisfy the following properties
    \begin{description}
        \item[Continuity] $\rho_{AB} \mapsto \entbA{A}{\rho_{AB}}{\sigma_{AB}}$ is continuous;
        \item[Non-negativity] $\entbA{A}{\rho_{AB}}{\sigma_{AB}} \ge 0$;
        \item[Semi-superadditivity] $\entbA{A}{\rho_{AB}}{\sigma_{A}\otimes \sigma_{B}} \ge \ent{\rho_A}{\sigma_A}$, and if $\rho_{AB}=\rho_A\otimes \rho_B$ then $\entbA{A}{\rho_{A}\otimes \rho_{B}}{\sigma_{A}\otimes \sigma_{B}} = \ent{\rho_A}{\sigma_A}$.
    \end{description}
     Moreover, $\entdA{A}{\cdot}{\cdot}$ is the only function $\mcl{S}(\hs_{AB}) \times \mcl{S}(\hs_{AB})\to \mathds{R}$ satisfying these three conditions plus the following:
    \begin{description}
        \item[Semi-monotonicity] For every CPTP map $T$, it holds that
        \begin{multline}
            \entdA{A}{T(\rho_{AB})}{T(\sigma_{AB})} + \entdA{B}{\Tr_A T(\rho_{AB})}{\Tr_A T(\sigma_{AB})} \\\le \entdA{A}{\rho_{AB}}{\sigma_{AB}} + \entdA{B}{\Tr_A \rho_{AB}}{\Tr_A \sigma_{AB}}.
        \end{multline}
    \end{description}
\end{proposition}
Non-negativity of $D^d_A$ is a classical result due to \cite{Lindblad1974,Lindblad1975, Uhlmann1977}, see also \cite{petz_monotonicity_2003, Vershynina2013}.

\subsection{Common zeros}
It is also interesting to compare the states over which the different definitions vanish. We start by showing some simple consequences of the definitions.
\begin{remark}
$\entA {\mcl{B}(\hs)|\vna} {\rho}{\sigma} =0$ if and only if $\rho = \condexp_\sigma^*(\rho)$, while $\entrA{\mcl{B}(\hs)|\vna}{\rho}{\sigma} = 0$ if and only if $\rho = \mcl{R}_\sigma(\rho)$.
With the notation of Subsection~\ref{sec:cond-exp-coarse-graining}, this implies that
\begin{equation}
\entrA{\mcl{B}(\hs)|\vna}{\rho}{\sigma} = 0 \iff \entA {\mcl{B}(\hs)|\mcl{F}_{\mcl{R}^*_\sigma}} {\rho}{\sigma} = 0.
 \label{eq:petz-expectation-fixed-points}
\end{equation}
\end{remark}
\begin{proposition}
\begin{equation}
\label{eq:ent-d-vs-fixed-point}
\entdA{\mcl{B}(\hs)|\vna}{\rho}{\sigma} \le \inf_{\substack{\mcl{M}\subset\vna \\ \sigma \in \mcl{S}(\mcl{M})}}
\entA {\mcl{B}(\hs)|\mcl{M}} {\rho}{\sigma} =
\entA {\mcl{B}(\hs)| \vna_\sigma} {\rho}{\sigma}.
\end{equation}
\end{proposition}
\begin{proof}
    First we observe that $D^d$ is monotone decreasing in the sub-algebra, in the sense that $\entdA{\mcl{B}(\hs)|\vna}{\rho}{\sigma} \le \entdA{\mcl{B}(\hs)|\mcl{M}}{\rho}{\sigma} $ whenever $\mcl{M}\subset \vna$. This is due to the data processing inequality since $\condexp_{\tau}^{\mcl{M}} \condexp_{\tau}^{\vna} = \condexp_{\tau}^{\mcl{M}}$. When $\mcl{M}$ is such that $\sigma \in \mcl{S}(\mcl{M})$, we have that $ \entdA{\mcl{B}(\hs)|\mcl{M}}{\rho}{\sigma} = \entA {\mcl{B}(\hs)|\mcl{M}} {\rho}{\sigma}$, proving the inequality. The equality follows from Lemma~\ref{lemma:accardi-cecchini-maximal}, showing that $\vna_\sigma$ is the largest subalgebra such that $\sigma \in \mcl{S}(\vna_\sigma)$.
\end{proof}

\begin{corollary}
    If $\entrA {\mcl{B}(\hs)|\vna} \rho \sigma = 0$, for some choice of  $\mcl{R}$, then $\entdA {\mcl{B}(\hs)|\vna} \rho \sigma =0$.
\end{corollary}
\begin{proof}
    If $\entrA {\mcl{B}(\hs)|\vna} \rho \sigma = 0$, then $\entA {\mcl{B}(\hs)|\mcl{F}_{\mcl{R}^*_\sigma}} {\rho}{\sigma} = 0$, hence the r.h.s. of \eqref{eq:ent-d-vs-fixed-point} vanishes.
\end{proof}
A converse of this result is provided by a classic result by Petz~\cite{petz_sufficient_1986}, which we reformulate in the specific setting we are presenting here.
\begin{theorem}[{Petz~\cite{petz_sufficient_1986}}]\label{thm:petz-sufficiency}
The following are equivalent:
\begin{enumerate}
    \item $\entdA {\mcl{B}(\hs)|\vna}{\rho}{\sigma} = 0$;
    \item\label{item:petz-modular-identity} $\rho^{z}\rho_\vna^{-z} =  \sigma^{z}\sigma_{\vna}^{-z}$  for all $z\in \mathds{C}$;
    \item $\rho^{1/2}\rho_\vna^{-1/2} =  \sigma^{1/2}\sigma_{\vna}^{-1/2}$.
\end{enumerate}
\end{theorem}
More precisely, in \cite{petz_sufficient_1986} condition~(\ref{item:petz-modular-identity}) of Theorem~\ref{thm:petz-sufficiency} is only shown to hold for $z$ in the strip $0 \le \Re z \le 1/2$ (which is the correct statement in the more general setting considered there), but if $\rho$ and $\sigma$ are full rank and finite dimensional the left and right hand sides of that identity are entire and therefore if they coincide on the strip, they coincide everywhere.
Note this has immediate implications for the Petz recovery maps and its variants, but not necessarily for other recovery maps included in Definition~\ref{def:recovery-map}.
\begin{proposition}\label{prop:common-kernel}
    If $\mcl{R}$ is one of $\mcl{R}^{P}$, $\mcl{R}^{P,t}$ for any $t\in \mathds{R}$, or $\mcl{R}^{PT}$, then we have that
    \begin{equation}
        \entrA {\mcl{B}(\hs)|\vna} \rho \sigma = 0 \quad \iff \quad \entdA {\mcl{B}(\hs)|\vna} \rho \sigma =0.
    \end{equation}
\end{proposition}

In this context we should also mention the results of \cite{carlen_recovery_2020}, in which they obtain the following ``quantitative'' version of Theorem \ref{thm:petz-sufficiency}:
\[
\entdA{\mcl{B}(\hs)|\vna}{\rho}{\sigma} \ge \qty(\frac{\pi}{4})^4 \norm{\Delta_{\sigma,\rho}}^{-2} \norm*{\sigma_\vna^{1/2}\rho_\vna^{-1/2}\rho^{1/2} -\sigma^{1/2}}_2^4 \ge
\qty(\frac{\pi}{8})^4 \norm{\Delta_{\sigma,\rho}}^{-2}
\norm*{\mcl{R}^P_\rho(\sigma_\vna) - \sigma}_1^4,
\]
where $\Delta_{\sigma,\rho}(x) = \sigma x \rho^{-1}$ is the relative modular operator.
(Note that in the r.h.s. of the last bound the roles of $\rho$ and $\sigma$ are exchanged).

\begin{remark}[Conditional Mutual Information]
A particularly interesting connection can be found when we take $\hs = \hs_{ABC} = \hs_A \otimes \hs_B \otimes \hs_C$, and $\vna = \mcl{B}(\hs_{AB}) \otimes \identity_{C}$. Then, for a fixed $\rho \in \mcl{S}(\hs_{ABC})$, choose $\sigma = \tau_A \otimes \rho_{BC} = \condexp^{BC}_{\tau}(\rho)$. We then have that \cite[Section~3.3]{CLP18b}:
\[
\entdA{C}{\rho}{\sigma} = I_\rho(A:C|B) = -S(\rho_{ABC}) +S(\rho_{BC}) + S(\rho_{AB}) - S(\rho_B),
\]
with $S(\omega)=-\Tr[\omega \log \omega]$ denoting the von Neumann entropy and $I_\rho(A:C|B)$ denoting the Conditional Mutual Information (CMI) \cite{fawzi_quantum_2015}. States $\rho$ for which $I_{\rho}(A:C|B) = 0$ are known as \emph{quantum Markov chains} in the  literature \cite{hayden_structure_2004}. Proposition~\ref{prop:common-kernel}, when specialized to this particular setup, is a restatement of the characterization of quantum Markov chains as states such that $\rho_{ABC} = \rho_{BC}^{1/2}\rho_{B}^{-1/2}\rho_{AB} \rho_{B}^{-1/2}\rho_{BC}^{1/2}$. By Theorem~\ref{thm:petz-sufficiency}, these are exactly the states for which $\rho_{ABC}^z\rho_{AB}^{-z}=\rho_{BC}^z\rho_{B}^{-z}$ for every $z\in\mathds{C}$. A weaker notion, known as \emph{Belavkin-Staszewski quantum Markov chains} \cite{bluhm_belavkin-staszewski_2025}, only requires this condition to be satisfied for $z=1$.
\end{remark}

\subsection{Relative bounds}\label{sec:relative-bounds}
Given that Proposition~\ref{prop:common-kernel} establishes that $D^d$ and $D^{\mcl{R}}$ vanish on the same set of states, when $\mcl{R}$ is the Petz recovery map or one of its rotated variants, a natural question to ask is if there is a simple bound between the two definitions of conditional relative entropy, in the sense of whether $\entdA{A}{\rho}{\sigma} \le \entrA{A}{\rho}{\sigma}$ or its opposite $\entdA{A}{\rho}{\sigma} \ge \entrA{A}{\rho}{\sigma}$ holds. It turns out that neither will hold in general.

Useful examples can be obtained via the connection to bounds on the CMI in the tripartite case.
Suppose that $\mcl{R}$ is an admissible recovery map satisfying $\mcl{R}_{\tau_A\otimes \rho_{BC}}=\operatorname{id}_A\otimes \mcl{R}_{\rho_{BC}}$ (such as the Petz recovery map or one of its variants). Then we have that
\[
\entdA{C}{\rho}{\tau_A \otimes \rho_{BC}} = I_\rho(A:C|B), \quad
\entrA{C}{\rho}{\tau_A \otimes \rho_{BC}} = \ent{\rho}{\mcl{R}_{\rho_{BC}}(\rho_{AB})}.
\]
Numerical counterexamples have been found to both $I_\rho(A:C|B) \le \ent{\rho}{\mcl{R}_{\rho_{BC}}(\rho_{AB})}$ and $I_\rho(A:C|B) \ge \ent{\rho}{\mcl{R}_{\rho_{BC}}(\rho_{AB})}$, for the case of the Petz recovery map \cite{brandao_quantum_2015, li_squashed_2018}. These were obtained for pure states, but a simple continuity argument shows that neither inequality can hold true for the case of full-rank states as well.

More generally, it has been shown \cite{fawzi_efficient_2018} that optimizing over any possible quantum channel $T:\mcl{B}(\hs_B)\to \mcl{B}(\hs_{BC})$ yields examples of states $\rho$ for which
\[
\min_{\substack{T \\ \text{quantum} \\\text{channel}}} \ent{\rho}{T(\rho_{AB})}
> I_\rho(A:C|B) = \entdA{C}{\rho}{\tau_A\otimes \rho_{BC}},
\]
which in particular implies that the inequality $D^d_C \ge D_C^{\mcl{R}}$ is false for this class of recovery maps.

\begin{remark}
The above discussion was considering a fixed recovery map, and asking for inequalities valid for any choice of $\rho$ and $\sigma$. If we instead fix $\sigma$, we can always find an admissible recovery map such that $\entrA{\mcl{B}(\hs)|\vna}{\rho}{\sigma} \ge \entdA{\mcl{B}(\hs)|\vna}{\rho}{\sigma}$ for every $\rho$, but the map will depend on $\sigma$ in general.
To see this, consider the replacement channel $T_\sigma(\rho) = \sigma \tr(\rho)$, then
\[
 \ent {\rho}{ T_\sigma(\rho)} = \ent \rho \sigma \ge \entdA{\mcl{B}(\hs)|\vna}{\rho}{\sigma},
\]
with strict inequality unless $\rho_\vna = \sigma_\vna$.
Note that $T_\sigma$ is not by itself an admissible recovery map, in the sense of Definition~\ref{def:recovery-map}, since it does not extend the dual conditional expectation $\condexp_\sigma^*$ on $\vna$, but it can be glued to it in a discontinuous way, i.e., by defining
\[
\mcl{R}(\sigma,\rho) = \begin{cases}
\condexp_\sigma^*(\rho), & \text{if } \sigma  \in \mcl{S}(\vna),\\
\sigma\tr \rho, & \text{otherwise.}
\end{cases}
\]
The discontinuity could also be avoided by smoothing it with a bump function. With this choice of recovery map, we then have that $\entrA{\mcl{B}(\hs)|\vna}{\rho}{\sigma} \ge \entdA{\mcl{B}(\hs)|\vna}{\rho}{\sigma}$ for every $\rho$ as claimed. Indeed, if $\sigma\in\mcl{S}(\vna)$, the two quantities coincide by \eqref{eq:cond-entropy-expectations}; otherwise, $\entrA{\mcl{B}(\hs)|\vna}{\rho}{\sigma}=\ent{\rho}{\sigma}$.
\end{remark}

\subsection{Other lower bounds}
\label{subsec:recovery-lower-bounds}
In the context of state recoverability theory, the conditional relative entropy by marginal differences is known as ``relative entropy difference''\footnote{Note that the definition of relative entropy difference for the case of Rényi entropy follows a very different pattern, which only coincides with the above definition for $\alpha=1$.}.
In this area, there are a number of results \cite{sutter_strengthened_2016, sutter_universal_2016, SBT17, junge_universal_2018} which provide lower bounds to $D^d_A$ in terms of a \emph{measured relative entropy}, in the sense that for certain choices of recovery map $\mcl{R}$ it holds that
\begin{equation}
\entdA{A}{\rho}{\sigma} \ge D_{\mathds{M}}(\rho_{AB}||{\mcl{R}}_{\sigma_{AB}}(\rho_{AB})),\quad
D_{\mathds{M}}(\rho||\sigma) = \sup_{\mcl{M}} H(\mcl{M}(\rho) || \mcl{M}(\sigma)),
\end{equation}
where the supremum is taken over all possible measurements (hence it is a classical, Kullback–Leibler divergence between the outcomes of the measurements). Interestingly, this lower bound has been proven for the averaged rotated Petz recovery map $\mcl{R}^{PT}$ \cite[Theorem 4.1]{SBT17}, but not for the regular Petz map.

A relaxation of the measured relative entropy is given in terms of quantum fidelity: if $F(\rho,\sigma) = \norm{\sqrt{\rho}\sqrt{\sigma}}_1^2$, then $D_{\mathds{M}}(\rho||\sigma) \ge -\log F(\rho,\sigma)$ (see \cite[Section~4]{SBT17}). It has been conjectured \cite{Seshadreesan2015} that the following lower bound holds for the Petz recovery map:
 \begin{equation}
\entdA{A}{\rho_{AB}}{\sigma_{AB}} \ge -\log F(\rho_{AB}, \mcl{R}^P_{\sigma_{AB}}(\rho_{AB})).
 \end{equation}
 This was disproved in \cite[Example~1]{Bhattacharya2026Recovery} by a qubit counterexample for diagonal pinching.\footnote{The failure also occurs for full-rank bipartite states under partial trace: embed each qubit state via $|i\rangle\mapsto|ii\rangle$, and replace each resulting state $\omega$ by $(1-\varepsilon)\omega+\varepsilon\identity_{AB}/4$ for sufficiently small $\varepsilon>0$.} Consequently, the stronger measured-relative-entropy bound above also fails in general for the standard Petz map. Valid recovery bounds using modified Petz maps are proved in \cite{berta_monotonicity_2015,Wilde2015,junge_universal_2018,SBT17}. Recovery corrections to data processing beyond finite-dimensional algebras are developed in \cite{JungeLaRacuente22Recovery}.

\section{Quasi-factorization of the quantum relative entropy}
\label{sec:quasi-factorization}

Quasi-factorization inequalities bound a global relative entropy by a sum of conditional relative entropies, with a multiplicative constant independent of the state $\rho$. Their usefulness comes from controlling this constant in terms of the reference state and the algebras on which we condition. We first consider the conditional relative entropy defined through a $\sigma$-preserving conditional expectation. We then discuss the extensions by marginal difference and by recovery, for which $\sigma$ need not admit such an expectation. All states below are full-rank.

\subsection{Conditional relative entropy}
\label{subsec:qf-conditional}

Let $\vna_1,\vna_2\subseteq\mcl{B}(\hs)$ be unital $*$-subalgebras, put $\mcl{M}=\vna_1\cap\vna_2$, and assume that
$\sigma\in\mcl{S}(\vna_1)\cap\mcl{S}(\vna_2) \subset \mcl{S}(\mcl{M})$. Let us denote the unique
$\sigma$-preserving conditional expectations by
$\condexp_j:=\condexp^{\vna_j}_\sigma$ and
$\condexp_0:=\condexp^{\mcl{M}}_\sigma$ respectively. Since they are $\sigma$-GNS-orthogonal projections onto the corresponding sub-algebra, they satisfy
$\condexp_j\condexp_0=\condexp_0\condexp_j=\condexp_0$.

\begin{definition}[Quasi-factorization]
\label{def:qf-conditional}
We say that  $\vna_1$ and $\vna_2$ satisfy \textit{quasi-factorization} if for every $\sigma\in\mcl{S}(\vna_1)\cap\mcl{S}(\vna_2)$ there is a constant $c=c(\sigma)\geq1$ such that
\begin{equation}\label{eq:qf-conditional}
 \entA{\mcl{B}(\hs)|\mcl{M}}{\rho}{\sigma}
 \leq c\left(\entA{\mcl{B}(\hs)|\vna_1}{\rho}{\sigma}
 +\entA{\mcl{B}(\hs)|\vna_2}{\rho}{\sigma}\right), \quad \forall \rho\in\mcl{S}(\hs).
\end{equation}
The case $c=1$ is called \textit{exact tensorization}.
\end{definition}

Quasi-factorization is also frequently called in the literature \emph{approximate tensorization of the relative entropy} \cite{CaputoMenzTetali15,BCR22,GaoRouze22}. This terminology refers to the classical tensorization property: for a product reference measure, the relative entropy is bounded by the sum of the single-site conditional relative entropies, with a multiplicative constant of one. Approximate tensorization extends this bound to dependent variables by allowing a multiplicative constant $c\geq1$; see \cite{CaputoMenzTetali15}. The same terminology is used for quantum conditional expectations in \cite{BCR22,GaoRouze22,LaRacuente22} and for quantum spin systems in \cite{capel_modified_2021}.

When $\mcl{M}=\mathds{C}\identity$, we have $\condexp_0^*(\rho)=\sigma$, so the left-hand side is simply $\ent{\rho}{\sigma}$. For a nontrivial intersection, the information retained in $\mcl{M}$ must remain as conditioning information: a state fixed by both dual expectations need not equal $\sigma$.

\subsubsection{Commuting squares and approximate commuting squares}

Let us begin with the easiest possible scenario, namely that of commuting squares of conditional expectations. The following entropy inequality is recalled in \cite{BCR22}.

\begin{proposition}[Exact tensorization for a commuting square]
\label{prop:qf-commuting-square}
If $\condexp_1\condexp_2=\condexp_2\condexp_1=\condexp_0$, then \eqref{eq:qf-conditional} holds with $c=1$.
\end{proposition}
\begin{proof}
It follows from the chain rule for conditional expectations \cite{ohya_quantum_2004} (see also \cite{JungeLaRacuenteRouze23}) and the data processing inequality, by
\begin{multline*}
\ent{\rho}{\condexp_0^*(\rho)}=\ent{\rho}{\condexp_1^*(\rho)}+\ent{\condexp_1^*(\rho)}{\condexp_0^*(\rho)} \\ = \ent{\rho}{\condexp_1^*(\rho)} + \ent{\condexp_1^*(\rho)}{\condexp_1^*\condexp_2^*(\rho)} \le \ent{\rho}{\condexp_1^*(\rho)} + \ent{\rho}{\condexp_2^*(\rho)}.
\end{multline*}
\end{proof}

\begin{example}
As an example, consider a multipartite system with product reference state, namely $\hs_\Lambda=\bigotimes_{x\in\Lambda}\hs_x$ and
$\sigma_\Lambda=\bigotimes_{x\in\Lambda}\sigma_x$, conditioning on the complement of $X\subseteq\Lambda$ gives
$\condexp_X^*(\rho_\Lambda)=\sigma_X\otimes\rho_{\Lambda\setminus X}$.
We write $D_X$ for the resulting conditional relative entropy, consistently with the bipartite notation. These expectations form commuting squares, and iteration yields
\begin{equation}\label{eq:qf-product}
 \ent{\rho_\Lambda}{\sigma_\Lambda}
 \leq\sum_{x\in\Lambda}\entA{x}{\rho_\Lambda}{\sigma_\Lambda}.
\end{equation}
This gives the product-state tensorization of \cite[Theorem 10]{CLP18b} in the conditional expectation setting.
\end{example}


Even when the commuting-square condition fails, quasi-factorization holds provided that the composition $\condexp_1\condexp_2$ is sufficiently close to $\condexp_0$ in the order on positive maps. Corollary~\ref{cor:qf-order} below makes this condition precise. We use the following comparison from \cite{GJLL23}.

\begin{lemma}[{\cite[Lemma 2.3]{GJLL23}}]\label{lem:qf-order-comparison}
Let $\condexp$ be a conditional expectation and $\Psi$ a unital completely positive map such that $\condexp\Psi=\condexp$ and
\[
 (1-\varepsilon)\condexp\leq\Psi\leq(1+\varepsilon)\condexp,
 \qquad 0\leq\varepsilon<1.
\]
The inequalities mean that the differences are positive maps. Then
\begin{equation}\label{eq:qf-order-comparison}
 \ent{\rho}{\Psi^*(\rho)}\geq q(\varepsilon)
 \ent{\rho}{\condexp^*(\rho)},\qquad
 q(\varepsilon)=\frac{1-\varepsilon}{1+\varepsilon}
 -\frac{\varepsilon}{(1-\varepsilon)(2\log2-1)}.
\end{equation}
In particular, $q(1/10)\geq1/2$.
\end{lemma}

For readability, we omit the proof of Lemma~\ref{lem:qf-order-comparison} and refer to \cite{GJLL23}. We focus instead on the following corollary, which provides the quasi-factorization estimate needed here.

\begin{corollary}\label{cor:qf-order}
If $(1-\varepsilon)\condexp_0\leq\condexp_1\condexp_2
\leq(1+\varepsilon)\condexp_0$ and $q(\varepsilon)>0$, then
\eqref{eq:qf-conditional} holds with $c=q(\varepsilon)^{-1}$.
In particular, $\varepsilon=1/10$ gives $c=2$.
\end{corollary}
\begin{proof}
Set $\Psi=\condexp_1\condexp_2$. Since $\Psi^*(\rho)=\condexp_2^*\condexp_1^*(\rho)$ is fixed by $\condexp_2^*$, the conditional-expectation chain rule and data processing give
\begin{align*}
 \ent{\rho}{\Psi^*(\rho)}
 &=\ent{\rho}{\condexp_2^*(\rho)}
   +\ent{\condexp_2^*(\rho)}{\condexp_2^*\condexp_1^*(\rho)}\\
 &\leq\ent{\rho}{\condexp_2^*(\rho)}
   +\ent{\rho}{\condexp_1^*(\rho)}.
\end{align*}
We conclude by applying Lemma~\ref{lem:qf-order-comparison} with $\condexp=\condexp_0$.
\end{proof}

If the comparison holds in completely positive order, the same argument applies after tensoring every map with the identity on an arbitrary auxiliary system. This gives a constant uniform in the auxiliary dimension.


The preceding criterion compares a single composition $\condexp_1\condexp_2$ with $\condexp_0$ on every positive operator, with a relative error $\varepsilon$ small enough that $q(\varepsilon)>0$. The resulting quasi-factorization constant depends only on this error. A spectral criterion instead measures how much the composition contracts observables orthogonal to the intersection algebra in a weighted Hilbert-space norm.

Write $L_2(\sigma)$ for $\mcl{B}(\hs)$ equipped with the $\sigma$-KMS inner product introduced above.

\begin{theorem}[{\cite[Theorem~5.1]{GaoRouze22}}]
\label{thm:qf-spectral}
Suppose that
\[
 \lambda=\norm{\condexp_1\condexp_2-\condexp_0:
 L_2(\sigma)\longrightarrow L_2(\sigma)}<1.
\]
There is a finite constant $c\geq1$ for which \eqref{eq:qf-conditional} holds. Moreover, a finite constant can be chosen uniformly after tensoring all three expectations with the identity on any auxiliary system.
\end{theorem}

Since both expectations act as the identity on $\mcl{M}$, the condition $\lambda<1$ expresses strict contraction on its orthogonal complement:
\[
 \norm{\condexp_1\condexp_2(X)}_{KMS,\sigma}
 \leq\lambda\norm{X}_{KMS,\sigma}
 \qquad\text{whenever }\condexp_0(X)=0.
\]
Thus, each application of $\condexp_1\condexp_2$ reduces the distance to $\condexp_0(X)$ by a factor of at most $\lambda$. Iterating gives, for every $X\in\mcl{B}(\hs)$ and $n\geq1$,
\[
 \norm{(\condexp_1\condexp_2)^n(X)-\condexp_0(X)}_{KMS,\sigma}
 \leq\lambda^n\norm{X-\condexp_0(X)}_{KMS,\sigma}.
\]
Since $\lambda<1$, the right-hand side tends to zero, and hence the repeated compositions converge to $\condexp_0$.  The estimates in \cite{GaoRouze22} control the quasi-factorization constant through $\lambda$ and a weighted subalgebra index, so the quantitative bound depends on more than the convergence rate. In finite dimensions, $\lambda<1$ follows from the fact that the two expectations are orthogonal projections whose ranges intersect in $\mcl{M}$.

For the dynamical application below, recall that a \emph{quantum Markov semigroup} is a norm-continuous family $(P_t)_{t\geq0}$ of unital, completely positive maps on $\mcl{B}(\hs)$ satisfying $P_0=\operatorname{id}$ and $P_{s+t}=P_s\circ P_t$ for all $s,t\geq0$. Its Hilbert--Schmidt adjoints are quantum channels describing the evolution of states, $\rho_t=P_t^*(\rho)$; a state $\sigma$ is invariant if $P_t^*(\sigma)=\sigma$ for every $t\geq0$. The semigroup is GNS-symmetric with respect to $\sigma$ if each $P_t$ is self-adjoint for the $\sigma$-GNS inner product.

For a GNS-symmetric semigroup with a faithful invariant state, let $\condexp=\lim_{t\to\infty}P_t$ be the conditional expectation onto its fixed-point algebra. We say that the semigroup satisfies a \emph{modified logarithmic Sobolev inequality} (MLSI) with rate $\alpha>0$ if
\[
 \ent{P_t^*(\rho)}{\condexp^*(\rho)}
 \leq e^{-\alpha t}\ent{\rho}{\condexp^*(\rho)}
 \qquad\text{for all }t\geq0,\quad\rho\in\mcl{S}(\hs).
\]
If the invariant state is unique, then $\condexp^*(\rho)=\sigma$. The inequality is \emph{complete} (CMLSI) if the same rate $\alpha$ works after replacing $P_t^*$ and $\condexp^*$ by $P_t^*\otimes\operatorname{id}_R$ and $\condexp^*\otimes\operatorname{id}_R$, for every finite-dimensional auxiliary system $R$ and all full-rank joint states \cite[Section~3]{GaoRouze22}.

In \cite[Theorem~3.3]{GaoRouze22}, the preceding criterion is used to show that, in finite dimensions, every GNS-symmetric quantum Markov semigroup with a faithful invariant state satisfies a complete modified logarithmic Sobolev inequality (although the constant might depend, in general, on the dimension).  For a growing family of system sizes, the obstacle lies in showing that $\lambda$ can be bounded away from 1 uniformly in system size, which requires a more delicate quantitative control of this parameter.

For more than two algebras, \cite{LaRacuente22} gives a complete quasi-factorization criterion using mixtures of compositions of conditional expectations and comparison in completely positive order. In particular, for any finite family of tracial conditional expectations $\condexp_1,\ldots,\condexp_m$ on $\mcl{B}(\hs)$, with $\condexp_0$ the tracial expectation onto the intersection of their ranges, there exists $c<\infty$ such that
\begin{equation}\label{eq:qf-finite-family}
 \ent{\rho}{\condexp_0^*(\rho)}
 \leq c\sum_{j=1}^m\ent{\rho}{\condexp_j^*(\rho)}.
\end{equation}
The constant can again be chosen independently of an auxiliary system. The more general criterion of \cite{LaRacuente22} also allows weighted expectations.
\subsubsection{Changing the reference state}

Whereas the preceding criteria establish quasi-factori\-zation for a fixed reference state, we now explain how an existing estimate can be transferred to a different reference state while keeping the subalgebras fixed.

A noncommutative change of measure transfers estimates for tracial expectations to expectations preserving $\sigma$ \cite{JungeLaRacuenteRouze23}. The argument of \cite[Proposition~2 and Appendix~B.1]{BCR22} gives the following consequence.

\begin{proposition}\label{prop:qf-change-measure}
Let $\condexp_j^{(0)}$ be the tracial expectations onto the same algebras as $\condexp_j$, for $j=0,1,2$.\footnote{The superscript $(0)$ labels the expectations associated with the maximally mixed reference state $\identity/\dim\hs$, whereas $\condexp_j$ preserves $\sigma$. Thus $\condexp_j^{(0)}$ is the map denoted above by $\condexp_\tau$, with the index $j$ specifying the target algebra.} If
\[
 \ent{\rho}{\condexp_0^{(0)*}(\rho)}
 \leq c_0\left(
 \ent{\rho}{\condexp_1^{(0)*}(\rho)}
 +\ent{\rho}{\condexp_2^{(0)*}(\rho)}\right)
\]
for every $\rho\in\mcl{S}(\hs)$, then \eqref{eq:qf-conditional} holds with
\[
 c=c_0\frac{\lambda_{\max}(\sigma)}{\lambda_{\min}(\sigma)}.
\]
\end{proposition}

The proof compares the conditional entropy functionals before and after conjugation by $\sigma^{1/2}$; their variational characterization follows from the conditional-expectation chain rule \cite{ohya_quantum_2004}. The same comparison applies to finite families. Note that its dependence on the condition number of $\sigma$ can prevent uniform estimates as the system size grows.

\subsubsection{From entropy contraction to quasi-factorization}

Whereas the preceding argument transfers an existing quasi-factorization estimate between reference states, the following criterion derives one directly from contraction of relative entropy under the average of the two expectations.

\begin{theorem}[{\cite[Theorem~4.1]{GaoWang26}}]
\label{thm:qf-from-contraction}
Set $\Phi=(\condexp_1+\condexp_2)/2$. If, for some $0<\eta<1$ and every $\rho\in\mcl{S}(\hs)$,
\begin{equation}\label{eq:qf-average-contraction}
 \ent{\Phi^*(\rho)}{\condexp_0^*(\rho)}
 \leq\eta\ent{\rho}{\condexp_0^*(\rho)},
\end{equation}
then \eqref{eq:qf-conditional} holds with
\[
 c=\frac{3}{2(1-\eta)}.
\]
If the contraction estimate holds uniformly after adjoining an auxiliary system, the same is true of the resulting quasi-factorization estimate.
\end{theorem}

We refer to \cite{GaoWang26} for the proof, which bounds the entropy lost in averaging the images under the two dual expectations using the quantum Jensen--Shannon divergence \cite{MajteyLambertiPrato05}, without introducing any additional dimension or subalgebra-index factor. The theorem is then leveraged to prove a modified logarithmic Sobolev inequality for detailed balanced Quantum Markov Semigroups in finite dimensions, going via a hypercontractivity estimate.
As a converse result to Theorem \ref{thm:qf-from-contraction}, the chain rule and convexity show that \eqref{eq:qf-conditional} implies \eqref{eq:qf-average-contraction} with $\eta=1-1/(2c)$.

\subsubsection{Classical Gibbs samplers acting on quantum states}

For the quantum embedding of classical heat-bath dynamics, \cite{BCR22} gives a more explicit estimate. Let $\sigma_\Lambda$ be the Gibbs state of a Hamiltonian diagonal in a product basis, and let $\condexp_X$ denote the stationary conditional expectation of the embedded dynamics on $X$. Its range $\vna_X$ is the fixed-point algebra of that dynamics; it need not be the full algebra of observables on $X^c$.

For $A,B\subseteq\Lambda$, let $\mu^\omega_{A\cup B}$ be the classical Gibbs measure with boundary configuration $\omega$, and let $\condexp_X^\omega$ denote the corresponding classical conditional expectations. Define
\begin{equation}\label{eq:qf-classical-embedding-constant}
 c_1=\sup_\omega\norm{
 \condexp_A^\omega\condexp_B^\omega-\condexp_{A\cup B}^\omega
 :L_1(\mu^\omega_{A\cup B})\longrightarrow L_\infty}.
\end{equation}
If $c_1<1$, then, for every quantum state $\rho_\Lambda$,
\begin{equation}\label{eq:qf-classical-embedding}
 \entA{\mcl{B}(\hs_\Lambda)|\vna_{A\cup B}}{\rho_\Lambda}{\sigma_\Lambda}
 \leq\frac{1}{1-c_1}\sum_{X\in\{A,B\}}
 \entA{\mcl{B}(\hs_\Lambda)|\vna_X}{\rho_\Lambda}{\sigma_\Lambda}.
\end{equation}
At inverse temperature $\beta=0$ these expectations form a commuting square. The finite-temperature estimate applies even when $\rho_\Lambda$ has coherences in the classical basis. The pinching argument in \cite{BCR22} removes the additive remainder in this setting; the corresponding result for general algebras belongs to weak quasi-factorization.

\subsubsection{Davies expectations for commuting spin chains}

For a finite-range, translation-invariant commuting interaction on a one-dimensional chain, \cite{BCGLPR24} proves
\begin{equation}\label{eq:qf-davies-chain}
 \ent{\rho_\Lambda}{\sigma_\Lambda}
 \leq C(1+\log|\Lambda|)\sum_{i\in\Lambda}
 \ent{\rho_\Lambda}{\condexp_{i,\sigma_\Lambda}^*(\rho_\Lambda)}.
\end{equation}
Here $\sigma_\Lambda$ is the Gibbs state at a fixed inverse temperature, and $\condexp_{i,\sigma_\Lambda}$ is the stationary expectation for the local Davies generator at $i$, with the couplings specified in that paper. The constant $C$ is independent of $|\Lambda|$. Its proof combines block factorization with estimates within each block. The conditioning algebras are the fixed-point algebras of the local generators; the terms in \eqref{eq:qf-davies-chain} are therefore conditional-expectation entropies.

\subsection{Conditional relative entropy by marginal difference}
\label{subsec:qf-marginal}

We now allow arbitrary (full-rank) reference states. For $\rho,\sigma\in\mcl{S}(\hs_\Lambda)$, recall that
\[
 \entdA{X}{\rho}{\sigma}
 =\ent{\rho}{\sigma}-\ent{\rho_{X^c}}{\sigma_{X^c}}.
\]
The results below use this definition directly and do not require a $\sigma$-preserving conditional expectation onto the complement algebra.

\subsubsection{Comparison with fixed point algebra conditional expectations}

For a subalgebra $\vna$ and a coarse graining map $\mcl{A}_\sigma$, \eqref{eq:ent-d-vs-fixed-point} gives us that
\begin{equation}\label{eq:petz-expectation-comparison}
 \entdA{\mcl{B}(\hs)|\vna}{\rho}{\sigma}
 \leq\entA{\mcl{B}(\hs)|\mcl{F}_{\mcl{A}_\sigma}}{\rho}{\sigma}.
\end{equation}
Consequently, a quasi-factorization bound by marginal differences also gives a bound by these conditional-expectation entropies. For spatial regions, a converse can be obtained by slightly enlarging the regions, provided that the expectations are local.

\begin{figure}[H]
 \centering
 \begin{tikzpicture}[x=.9cm,y=1cm]
  \fill[figConditioning] (0,0) rectangle (12,.75);
  \draw[fill=figRegionB] (2,0) rectangle (10,.75);
  \draw[fill=figRegionA] (4,0) rectangle (8,.75);
  \draw (0,0) rectangle (12,.75);
  \node at (6,.375) {$A$};
  \node at (1,.375) {$A_\partial^c$};
  \node at (11,.375) {$A_\partial^c$};
  \foreach \x in {2,4,8,10} {
   \draw[thin] (\x,.75) -- (\x,1.18);
  }
  \draw[<->] (2,1.05) -- node[above] {$r$} (4,1.05);
  \draw[<->] (8,1.05) -- node[above] {$r$} (10,1.05);
  \node[anchor=south east] at (12,1.1) {$\Lambda$};
  \draw (2,-.12) -- (2,-.3) -- (10,-.3) -- (10,-.12);
  \node at (6,-.55) {$A_\partial$};
 \end{tikzpicture}
 \caption{Enlarging a block by the interaction range, illustrated on a chain. The blue block $A$ and the gold collars of width $r$ form $A_\partial$. The local Davies expectation associated with $A$ acts trivially on the purple complement $A_\partial^c$.}
 \label{fig:conditional-entropy-enlargement}
\end{figure}
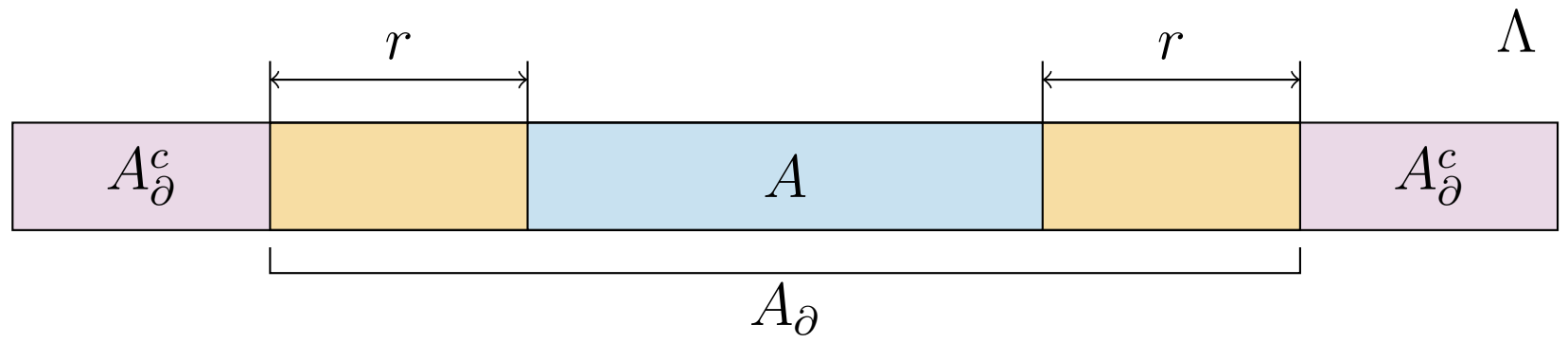

For example, consider the local Davies expectations for a Gibbs state $\sigma$ of a finite-range commuting Hamiltonian on a finite lattice $\Lambda$, with the single-site couplings of \cite[Section~5.1]{BCR22}. Let $\mcl{F}_A$ be the fixed-point algebra for the dynamics on a region $A\subseteq\Lambda$. Its stationary expectation $\condexp_\sigma^{\mcl{F}_A}$ acts nontrivially only on $A_\partial:=\{x\in\Lambda:\operatorname{dist}(x,A)\leq r\}$, where $r$ bounds the interaction range (see \cite[Section~5.1]{BCR22} and \cite[Appendix~A.3]{capel_quasi-optimal_2025}). This enlargement is illustrated in Figure~\ref{fig:conditional-entropy-enlargement}. In this setting,
\begin{equation}\label{eq:cre-comparison-enlarged-region}
 \entdA{A}{\rho}{\sigma}
 \leq\entA{\mcl{B}(\hs)|\mcl{F}_A}{\rho}{\sigma}
 \leq\entdA{A_\partial}{\rho}{\sigma}.
\end{equation}
The first inequality is \eqref{eq:petz-expectation-comparison}. For the second, locality implies that $\bigl(\condexp_\sigma^{\mcl{F}_A}\bigr)^*(\rho)$ has the same marginal as $\rho$ on $A_\partial^c$. The conditional-expectation chain rule \cite{ohya_quantum_2004} and data processing therefore give
\begin{align*}
 \entA{\mcl{B}(\hs)|\mcl{F}_A}{\rho}{\sigma}
 &=\ent{\rho}{\sigma}
 -\ent{\bigl(\condexp_\sigma^{\mcl{F}_A}\bigr)^*(\rho)}{\sigma}\\
 &\leq\ent{\rho}{\sigma}-\ent{\rho_{A_\partial^c}}{\sigma_{A_\partial^c}}
 =\entdA{A_\partial}{\rho}{\sigma}.
\end{align*}
Thus a quasi-factorization bound for the global relative entropy in terms of Davies conditional relative entropies on blocks $A_i$ gives one by marginal differences on the enlarged blocks $(A_i)_\partial$, with the same constant.

\subsubsection{General superadditivity and overlapping regions}

We first consider a bipartite Hil\-bert space $\hs_{AC}=\hs_A\otimes\hs_C$ and states $\rho_{AC},\sigma_{AC}\in\mcl{S}(\hs_{AC})$.  The starting point is the superadditivity for general states from \cite[Theorem~1]{CLP18}. Set
\begin{equation}\label{eq:qf-correlation-operator}
 H(\sigma_{AC}):=
 (\sigma_A^{-1/2}\otimes\sigma_C^{-1/2})\sigma_{AC}
 (\sigma_A^{-1/2}\otimes\sigma_C^{-1/2})-\identity_{AC}.
\end{equation}
The operator $H(\sigma_{AC})$ measures the deviation of $\sigma_{AC}$ from the product of its marginals in a relative operator sense. In particular, it vanishes if and only if $\sigma_{AC}=\sigma_A\otimes\sigma_C$.

The general superadditivity bound reads
\begin{equation}\label{eq:qf-superadditivity}
 \ent{\rho_A}{\sigma_A}+\ent{\rho_C}{\sigma_C}
 \leq(1+2\norm{H(\sigma_{AC})}_\infty)
 \ent{\rho_{AC}}{\sigma_{AC}}.
\end{equation}
Note that for a product state $\sigma_{AC}$, inequality~\eqref{eq:qf-superadditivity} recovers the standard superadditivity of the relative entropy, and otherwise up to a factor depending only on how far $\sigma_{AC}$ is from being a product state. The estimate improves on the factor two obtained directly from data processing whenever $\norm{H(\sigma_{AC})}_\infty<1/2$.

Inequality~\eqref{eq:qf-superadditivity} is equivalent to the following formulation as a quasi-factorization of the relative entropy, for which we give a short proof.

\begin{theorem}[{\cite[Theorem~5]{CLP18b}}]
\label{thm:qfAB}
Let $\rho_{ABC},\sigma_{ABC}\in\mcl{S}(\hs_A\otimes\hs_B\otimes\hs_C)$. Then
\begin{equation}\label{eq:qfAB}
 (1-2\norm{H(\sigma_{AC})}_\infty)\ent{\rho_{ABC}}{\sigma_{ABC}}
 \leq\entdA{AB}{\rho_{ABC}}{\sigma_{ABC}}
     +\entdA{BC}{\rho_{ABC}}{\sigma_{ABC}}.
\end{equation}
\end{theorem}
\begin{proof}
By data processing \cite{Lindblad1974,Uhlmann1977},
\begin{align*}
 &\ent{\rho_{ABC}}{\sigma_{ABC}}
   -\ent{\rho_A}{\sigma_A}-\ent{\rho_C}{\sigma_C}\\
 &\qquad\qquad\qquad\geq\ent{\rho_{AC}}{\sigma_{AC}}
   -\ent{\rho_A}{\sigma_A}-\ent{\rho_C}{\sigma_C}\\
 &\qquad\qquad\qquad=\Tr\Bigl[\rho_{AC}\Bigl(\log\rho_{AC}-\log\underbrace{
 \exp\bigl(\log\sigma_{AC}-\log(\sigma_A\otimes\sigma_C)
       +\log(\rho_A\otimes\rho_C)\bigr)}_{=:M}\Bigr)\Bigr]\\
 &\qquad\qquad\qquad\geq-\log\Tr M,
\end{align*}
where the last inequality follows from non-negativity of $\ent{\rho_{AC}}{M/\Tr M}$.
To bound the last term, define
\begin{align*}
 \mcl{T}_{\sigma_A\otimes\sigma_C}(X)
 &=\int_0^\infty(\sigma_A\otimes\sigma_C+s\identity)^{-1}
 X(\sigma_A\otimes\sigma_C+s\identity)^{-1}\,ds,\\
 L&=\mcl{T}_{\sigma_A\otimes\sigma_C}(\sigma_{AC})-\identity.
\end{align*}
Lieb's three-matrix trace inequality \cite[Theorem~7]{Lieb1973} yields
\[
 \Tr M\leq\Tr\bigl[(\rho_A\otimes\rho_C)
 \mcl{T}_{\sigma_A\otimes\sigma_C}(\sigma_{AC})\bigr]
 =1+\Tr[ L(\rho_A\otimes\rho_C)].
\]
Note that $\norm{L}_\infty\leq\norm{H(\sigma_{AC})}_\infty$.
This follows from the fact that
\[
 -\norm{H(\sigma_{AC})}_\infty(\sigma_A\otimes\sigma_C)
 \leq\sigma_{AC}-\sigma_A\otimes\sigma_C
 \leq\norm{H(\sigma_{AC})}_\infty(\sigma_A\otimes\sigma_C),
\]
applying the positive map $\mcl{T}_{\sigma_A\otimes\sigma_C}$ to this inequality, and using
$\mcl{T}_{\sigma_A\otimes\sigma_C}(\sigma_A\otimes\sigma_C)=\identity$.
Moreover, the marginal identities give
$\Tr [L(X_A\otimes\sigma_C)]=\Tr [L(\sigma_A\otimes X_C)]=0$
for all $X_A,X_C$; this is the centering argument of \cite[Lemma~5]{CLP18}.
Thus, using $\log x\leq x-1$,
\begin{align*}
 \log\Tr M
 &\leq\Tr [L(\rho_A\otimes\rho_C)]
 =\Tr \bigl[L(\rho_A-\sigma_A)\otimes(\rho_C-\sigma_C)\bigr]\\
 &\leq\norm{H(\sigma_{AC})}_\infty
       \norm{\rho_A-\sigma_A}_1\norm{\rho_C-\sigma_C}_1\\
 &\leq2\norm{H(\sigma_{AC})}_\infty
       \sqrt{\ent{\rho_A}{\sigma_A}\ent{\rho_C}{\sigma_C}}\\
 &\leq2\norm{H(\sigma_{AC})}_\infty\ent{\rho_{ABC}}{\sigma_{ABC}}.
\end{align*}
Here we used the quantum Pinsker inequality and data processing in the last two steps. Consequently,
\begin{align*}
 \entdA{AB}{\rho_{ABC}}{\sigma_{ABC}}
 +\entdA{BC}{\rho_{ABC}}{\sigma_{ABC}}
 &=2\ent{\rho_{ABC}}{\sigma_{ABC}}
   -\ent{\rho_A}{\sigma_A}-\ent{\rho_C}{\sigma_C}\\
 &\geq(1-2\norm{H(\sigma_{AC})}_\infty)
   \ent{\rho_{ABC}}{\sigma_{ABC}},
\end{align*}
as claimed.
\end{proof}

A finite quasi-factorization constant follows when $2\norm{H(\sigma_{AC})}_\infty<1$. If $\sigma_{AC}=\sigma_A\otimes\sigma_C$, the constant is one, even if $\sigma_{ABC}$ is correlated with $B$. This is a condition on the two outer marginals, and does not assert that the conditional expectations onto $A$ or $C$ exist for $\sigma_{ABC}$.

\subsubsection{Multipartite marginal terms}

The same centering argument extends to any number of subsystems; this is a direct extension of the proof of \cite{CLP18}.

\begin{proposition}\label{prop:qf-multipartite-marginals}
Let $n\geq2$, $\Lambda=A_1\cdots A_n$, and
$\rho,\sigma\in\mcl{S}(\bigotimes_{i=1}^n\hs_{A_i})$. Put
\begin{align*}
 \kappa_n&=\norm{\left(\bigotimes_{i=1}^n\sigma_{A_i}\right)^{-1/2}
 \sigma\left(\bigotimes_{i=1}^n\sigma_{A_i}\right)^{-1/2}-\identity}_\infty,
 &a_n&=2\sum_{k=1}^{n-1}\sqrt{k}.
\end{align*}
Then
\begin{equation}\label{eq:qf-multipartite-superadditivity}
 \sum_{i=1}^n\ent{\rho_{A_i}}{\sigma_{A_i}}
 \leq\min\{n,1+a_n\kappa_n\}\ent{\rho}{\sigma}.
\end{equation}
Consequently,
\begin{equation}\label{eq:qf-multipartite-conditional}
 (n-1-a_n\kappa_n)\ent{\rho}{\sigma}
 \leq\sum_{i=1}^n\entdA{\Lambda\setminus A_i}{\rho}{\sigma}.
\end{equation}
\end{proposition}
\begin{proof}
The argument of Theorem~\ref{thm:qfAB}, applied to the product of $n$ marginals, gives
\[
 \sum_{i=1}^n\ent{\rho_{A_i}}{\sigma_{A_i}}-\ent{\rho}{\sigma}
 \leq\Tr\left[L\left(\bigotimes_{i=1}^n\rho_{A_i}\right)\right],
\]
where $L=\mcl{T}_{\bigotimes_{i=1}^n\sigma_{A_i}}(\sigma)-\identity$
and $\norm{L}_\infty\leq\kappa_n$ by the same positivity argument.
Only the centering estimate changes. The marginal identities now give
\[
 \Tr\left[L\left(\bigotimes_{j<i}\sigma_{A_j}\otimes X_{A_i}
              \otimes\bigotimes_{j>i}\sigma_{A_j}\right)\right]=0
\]
for every $X_{A_i}$ and every $i$. Expanding the product by telescoping and cancelling the terms with at most one factor $\rho_{A_i}-\sigma_{A_i}$ therefore yields
\begin{align*}
 \Tr\left[L\left(\bigotimes_{i=1}^n\rho_{A_i}\right)\right]
 &\leq\kappa_n\sum_{k=1}^{n-1}
 \norm{\bigotimes_{i=1}^k\rho_{A_i}-\bigotimes_{i=1}^k\sigma_{A_i}}_1
 \norm{\rho_{A_{k+1}}-\sigma_{A_{k+1}}}_1\\
 &\quad\leq2\kappa_n\sum_{k=1}^{n-1}\sqrt{k}\,\ent{\rho}{\sigma}
 =a_n\kappa_n\ent{\rho}{\sigma}.
\end{align*}
For the last inequality, Pinsker's inequality, additivity on product states and data processing yield
\begin{align*}
 \norm{\bigotimes_{i=1}^k\rho_{A_i}-\bigotimes_{i=1}^k\sigma_{A_i}}_1
 &\leq\sqrt{2\sum_{i=1}^k\ent{\rho_{A_i}}{\sigma_{A_i}}}
 \leq\sqrt{2k\ent{\rho}{\sigma}},\\
 \norm{\rho_{A_{k+1}}-\sigma_{A_{k+1}}}_1
 &\leq\sqrt{2\ent{\rho}{\sigma}}.
\end{align*}
This proves the bound with coefficient $1+a_n\kappa_n$, meanwhile straightforward data processing gives the alternative coefficient $n$.
The conditional bound~\eqref{eq:qf-multipartite-conditional} follows from the definition of $D^d$, as in Theorem~\ref{thm:qfAB}.
\end{proof}

In general, the coefficient in \eqref{eq:qf-multipartite-conditional} is positive when $\kappa_n<(n-1)/a_n$.
At a product reference state, $\kappa_n=0$ and the coefficient is $n-1$, reflecting that each subsystem belongs to $n-1$ of the complementary regions.

A direct application is to a channel that chooses $i$ uniformly, retains the marginal on $A_i$, and reconstructs the full state using the Petz recovery map for $\sigma$. Denote the resulting averaged channel by $\Phi_\sigma$, so that $\Phi_\sigma(\sigma)=\sigma$. Joint convexity, data processing for each recovery map, and \eqref{eq:qf-multipartite-superadditivity} give
\[
 \ent{\Phi_\sigma(\rho)}{\sigma}
 \leq\frac1n\sum_{i=1}^n\ent{\rho_{A_i}}{\sigma_{A_i}}
 \leq q\,\ent{\rho}{\sigma},
 \qquad q=\frac{1+a_n\kappa_n}{n}.
\]
Thus, whenever $\kappa_n<(n-1)/a_n$, iterating this channel yields
$\ent{\Phi_\sigma^m(\rho)}{\sigma}\leq q^m\ent{\rho}{\sigma}$ for every $m\geq1$, giving an explicit exponential rate of convergence to $\sigma$.

\subsubsection{Exact factorization for Markov states}

We now consider quasi-factorization inequalities in which the left-hand side is itself a conditional relative entropy by marginal difference, rather than the global relative entropy. These give a stronger formulation in the sense that factorization can be performed within a region while keeping its complement as a conditioning system; the global formulation is recovered when this complement is empty.

The Markov property of a state gives an exact factorization of conditional relative entropy by marginal differences. The following formulation follows from  \cite{BCLPR21}.

\begin{proposition}\label{prop:qf-markov-reference}
Let $\rho_{ABC},\sigma_{ABC}\in\mcl{S}(\hs_A\otimes\hs_B\otimes\hs_C)$ and suppose that $I_\sigma(A:C\mid B)=0$. Then
\begin{equation}\label{eq:qf-markov-identity}
 \begin{split}
 \entdA{A}{\rho_{ABC}}{\sigma_{ABC}}
 +\entdA{C}{\rho_{ABC}}{\sigma_{ABC}}
 -\entdA{AC}{\rho_{ABC}}{\sigma_{ABC}}=I_\rho(A:C\mid B).
 \end{split}
\end{equation}
In particular,
\begin{equation}\label{eq:qf-markov-reference}
 \entdA{AC}{\rho_{ABC}}{\sigma_{ABC}}
 \leq\entdA{A}{\rho_{ABC}}{\sigma_{ABC}}
 +\entdA{C}{\rho_{ABC}}{\sigma_{ABC}}.
\end{equation}
\end{proposition}
\begin{proof}
 Expanding the marginal differences gives
\begin{align*}
 \entdA{A}{\rho_{ABC}}{\sigma_{ABC}}
& +\entdA{C}{\rho_{ABC}}{\sigma_{ABC}}
 -\entdA{AC}{\rho_{ABC}}{\sigma_{ABC}}\\
 &=\ent{\rho_{ABC}}{\sigma_{ABC}}+\ent{\rho_B}{\sigma_B}
 -\ent{\rho_{AB}}{\sigma_{AB}}-\ent{\rho_{BC}}{\sigma_{BC}}\\
 &=I_\rho(A:C\mid B)
 -\Tr\left[\rho_{ABC}\left(
 \log\sigma_{ABC}+\log\sigma_B-\log\sigma_{AB}-\log\sigma_{BC}
 \right)\right].
\end{align*}
The logarithmic expression vanishes by the Markov property of $\sigma_{ABC}$ \cite[Proposition~5]{BCLPR21}. Strong subadditivity gives the inequality.
\end{proof}

\begin{figure}[htbp]
 \centering
 \begin{tikzpicture}[x=1cm,y=1cm]
  \fill[figConditioning] (0,0) rectangle (10.5,.65);
  \foreach \x/\i in {1/1,4.25/2,7.5/3} {
   \draw[fill=figRegionA] (\x,0) rectangle (\x+2,.65);
   \node at (\x+1,.325) {$A_{\i}$};
  }
  \draw (0,0) rectangle (10.5,.65);
  \node at (5.25,1) {$\Lambda$};
 \end{tikzpicture}
 \caption{Separated blocks, illustrated for $A=A_1\cup A_2\cup A_3$. The regions outside the labelled blocks together form the conditioning system $A^c$ in the Markov conditions for \eqref{eq:qf-markov-blocks}.}
 \label{fig:qf-markov-blocks}
\end{figure}
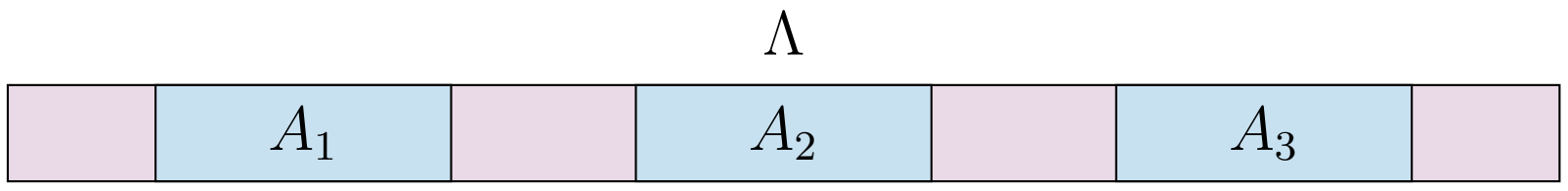

Iterating this argument over separated regions gives the block factorization used in \cite{BCLPR21}. Explicitly, let $A_1,\ldots,A_n\subseteq\Lambda$ be pairwise disjoint as in Figure \ref{fig:qf-markov-blocks}, set $A=\bigcup_{i=1}^n A_i$, and suppose that $I_\sigma(A_i:A\setminus A_i\mid A^c)=0$ for every $i$. Then, for every $\rho_\Lambda\in\mcl{S}(\hs_\Lambda)$,
\begin{equation}\label{eq:qf-markov-blocks}
 \entdA{A}{\rho_\Lambda}{\sigma_\Lambda}
 \leq\sum_{i=1}^n\entdA{A_i}{\rho_\Lambda}{\sigma_\Lambda}.
\end{equation}

\subsubsection{Operator lower bounds for general  states}

The preceding exact factorization relies on the Markov property of the state; operator lower bounds provide a quasi-factorization estimate for general faithful  states, at the cost of a multiplicative constant.

On $\hs_\Lambda=\bigotimes_{i\in\Lambda}\mathds{C}^d$, let $\vna_V=\identity_V\otimes\mcl{B}(\hs_{V^c})$ for $V\subseteq\Lambda$. Recall that the tracial conditional expectation onto $\vna_V$ is
\[
 \condexp_\tau^{\vna_V}(Z)=\frac{\identity_V}{d^{|V|}}\otimes\Tr_V Z,
 \qquad \tau=\frac{\identity_\Lambda}{d^{|\Lambda|}}.
\]
This map replaces the state on $V$ by the maximally mixed state.

The following result converts lower bounds on $\sigma$ relative to its images under these tracial expectations into a bound on the conditional relative entropy on $A\cup B$ by the sum of those on $A$ and $B$.

\begin{proposition}[{\cite[Lemma~E.4]{GaoGuoWang26}}]
\label{prop:qf-reference-lower-bounds}
Let $\rho,\sigma\in\mcl{S}(\hs_\Lambda)$ and let $A,B\subseteq\Lambda$ be nonempty, possibly overlapping. Suppose
\[
 \sigma\geq m_A\condexp_\tau^{\vna_A}(\sigma),\qquad
 \sigma\geq m_B\condexp_\tau^{\vna_B}(\sigma),\qquad 0<m_A,m_B\leq1.
\]
Then
\begin{equation}\label{eq:qf-reference-lower-bounds}
 \entdA{A\cup B}{\rho}{\sigma}
 \leq C_{A,B}\bigl(\entdA{A}{\rho}{\sigma}+\entdA{B}{\rho}{\sigma}\bigr),
\end{equation}
where
\[
 C_{A,B}=\frac{4d^{2|B|}}{m_B}
 \left(1+\frac{2(|A|+|B|)\log d-\log(m_Am_B)}{1-m_B/2}\right)
 +\frac{5}{1-m_B/2}.
\]
\end{proposition}
The proof is given in \cite[Appendix~E.1]{GaoGuoWang26}.
Every faithful $\sigma$ admits the required lower bounds with
$
 m_V={\lambda_{\min}(\sigma)}/{\norm{\condexp_\tau^{\vna_V}(\sigma)}_\infty},
 \text{ for } V=A,B. $
If $A\cup B=\Lambda$, the left-hand side of \eqref{eq:qf-reference-lower-bounds} is $\ent{\rho}{\sigma}$. Thus the estimate provides a finite quasi-factorization constant even when the correlation bound in Theorem~\ref{thm:qfAB} is not small. Its constant, however, need not remain bounded as the system grows. In \cite{GaoGuoWang26}, this quasi-factorization result is then used to show a spatial factorization of the entropy, leading to a system-size-independent complete modified logarithmic Sobolev inequality for the Gaussian and Metropolis Chen–Kastoryano–Gilyén Gibbs samplers on one-dimensional chains \cite{chen_efficient_2025}.

\subsubsection{Block factorization for commuting one-dimensional Gibbs states}

The next result uses rapid decay of correlations of $\sigma$ to obtain a quasi-factorization constant independent of the system size. We now assume that $\sigma$ is the Gibbs state of a one-dimensional, finite-range, translation-invariant, commuting interaction, for which this decay is available. Together with the Markov property, it yields a uniform estimate on overlapping blocks.

\begin{proposition}[{\cite[Section~3.1 and Lemma~3.3]{BCGLPR24}}]
\label{prop:qf-gibbs-blocks}
Let $\Lambda$ be a finite interval in $\mathbb Z$, with local dimension $d$, and let $H$ be the Hamiltonian of a finite-range, translation-invariant, commuting interaction. Its Gibbs state at a fixed inverse temperature $\beta>0$ is
\[
 \sigma=\frac{e^{-\beta H}}{\Tr[e^{-\beta H}]}.
\]
There is a cover
\[
 \Lambda=\bigcup_{i=1}^m(A_i\cup B_i)
\]
by two families of intervals, disjoint within each family and overlapping between families, whose lengths and overlaps are $O(\log|\Lambda|)$, such that
\begin{equation}\label{eq:qf-gibbs-blocks}
 \ent{\rho}{\sigma}
 \leq C\sum_{i=1}^m\left(
 \entdA{A_i}{\rho}{\sigma}
 +\entdA{B_i}{\rho}{\sigma}\right)
\end{equation}
for every $\rho\in\mcl{S}(\hs)$. The constant $C$ depends on the interaction, $\beta$ and $d$, but not on $|\Lambda|$.
\end{proposition}

The proof in \cite{BCGLPR24} first applies Theorem~\ref{thm:qfAB} to the unions of the two families. The Markov property then splits the resulting conditional terms into individual blocks, as in Proposition~\ref{prop:qf-markov-reference}. Logarithmic overlaps control the accumulated correlation error.

To pass to the usual conditional relative entropy, let $\mcl{F}_X$ be the fixed-point algebra of the local Davies generator on a block $X$, with the couplings considered in \cite{BCGLPR24}, and let $\condexp_\sigma^{\mcl{F}_X}$ be its stationary conditional expectation. The comparison \eqref{eq:petz-expectation-comparison}, proved in \cite[Proposition~7]{BCR22}, takes the explicit form
\begin{equation}\label{eq:qf-davies-block-comparison}
 \entdA{X}{\rho}{\sigma}
 \leq\ent{\rho}{\bigl(\condexp_\sigma^{\mcl{F}_X}\bigr)^*(\rho)}
 =\entA{\mcl{B}(\hs)|\mcl{F}_X}{\rho}{\sigma}.
\end{equation}
Combining this comparison with \eqref{eq:qf-gibbs-blocks} yields quasi-factorization for the usual conditional relative entropy, with the same constant $C$ \cite[Lemma~3.4]{BCGLPR24}:
\begin{equation}\label{eq:qf-davies-blocks}
 \ent{\rho}{\sigma}
 \leq C\sum_{i=1}^m\left(
 \entA{\mcl{B}(\hs)|\mcl{F}_{A_i}}{\rho}{\sigma}
 +\entA{\mcl{B}(\hs)|\mcl{F}_{B_i}}{\rho}{\sigma}\right).
\end{equation}
Here the conditioning algebras are the Davies fixed-point algebras. Further estimates within each block yield the single-site bound already stated in \eqref{eq:qf-davies-chain}. Combining these results, \cite{BCGLPR24} obtains a modified logarithmic Sobolev inequality for Quantum Markov Semigroups on a 1D chain, with constant scaling as inverse logarithm of the system size (a scaling that, while not being uniformly bounded away from zero, still allows to obtain the usual mixing time and convergence rates estimates for the semigroup).

\subsubsection{Uniform quasi-factorization for one-dimensional Gibbs states}

The conditional te\-rms in Proposition~\ref{prop:qf-multipartite-marginals} concern complements of individual subsystems. Proposition~\ref{prop:qf-gibbs-blocks} instead uses blocks whose size grows with the chain. The following result gives a uniform bound by single-site conditional terms for one-dimensional Gibbs states, without assuming that the interactions commute.

\begin{theorem}[{\cite[Theorem~3.1]{GaoGuoWang26}}]
\label{thm:qf-gibbs-1d}
Let $\Lambda$ be a finite chain or cycle with local dimension $d$, and let $H_\Lambda=\sum_{X\subseteq\Lambda}h_X$, where $h_X=h_X^*$ is supported on $X$ and
\[
 h_X=0\quad\text{if }\operatorname{diam}(X)>r_0,
 \qquad\sup_{i\in\Lambda}\sum_{X\ni i}\norm{h_X}_\infty\leq J_0.
\]
Let $\sigma_\Lambda$ be the Gibbs state of $H_\Lambda$ at inverse temperature $\beta>0$. There exists $C_s=C_s(\beta,d,r_0,J_0)$, independent of $|\Lambda|$, such that
\begin{equation}\label{eq:qf-gibbs-1d}
 \ent{\rho_\Lambda}{\sigma_\Lambda}
 \leq C_s\sum_{i\in\Lambda}\entdA{\{i\}}{\rho_\Lambda}{\sigma_\Lambda}
\end{equation}
for every $\rho_\Lambda\in\mcl{S}(\hs_\Lambda)$. More generally, for every finite reference system $R$ and every $\rho_{R\Lambda}\in\mcl{S}(\hs_R\otimes\hs_\Lambda)$,
\[
 \ent{\rho_{R\Lambda}}{\rho_R\otimes\sigma_\Lambda}
 \leq C_s\sum_{i\in\Lambda}
 \entdA{\{i\}}{\rho_{R\Lambda}}{\rho_R\otimes\sigma_\Lambda},
\]
with the same constant, independent of $\dim\hs_R$. Each partial trace retains $R$.
\end{theorem}
The interactions need not commute or be translation invariant. We refer to \cite[Appendix~G.2]{GaoGuoWang26} for the proof. The uniform constant uses one-dimensional Gibbs locality in addition to operator lower bounds.

\subsection{Conditional relative entropies by recovery}
\label{subsec:qf-recovery}

We prove here quasi-factorization results for general admissible recovery maps satisfying Definition~\ref{def:recovery-map}, using its axioms as the only assumptions on the maps. We first establish a finite-dimensional existence result, and then obtain a coefficient controlled by the interaction between two recovery maps. The latter argument is inspired by the quasi-factorization estimate for Petz recovery in \cite[Theorem~7 and Section~4.3]{CLP18b}, whose trace-inequality strategy we adapt to the axiomatic setting. We subsequently recall that paper's explicit estimate for Petz recovery. Throughout, the maps include the restriction to the conditioning algebra, as in our definition of $D^{\mcl{R}}$.

\subsubsection{Finite-dimensional factorization for a cover}

The following proposition is proved here by combining the recovery axioms with Petz's equality conditions for data processing \cite{petz_sufficient_1986} and elementary norm estimates. Its use of a positive quadratic gap to control relative entropy is analogous to the spectral approach to quasi-factorization for conditional expectations in \cite[Theorem~5.1]{GaoRouze22}. Here the gap measures the joint deviation from the fixed points of the recovery maps. The resulting constant is finite for each fixed faithful state, but is not asserted to be uniform in the system size.

\begin{proposition}\label{prop:qf-recovery-finite}
Let $\hs=\bigotimes_{x\in\Lambda}\hs_x$ have dimension at least two, let $\sigma\in\mcl{S}(\hs)$, and let $A_1,\ldots,A_m\subseteq\Lambda$ satisfy $\bigcup_{j=1}^mA_j=\Lambda$. For each $j$, let $\mcl{R}_j$ satisfy Definition~\ref{def:recovery-map} for $\identity_{A_j}\otimes\mcl{B}(\hs_{A_j^c})$. Write $\mcl{R}=(\mcl{R}_1,\ldots,\mcl{R}_m)$ and $\mcl{R}_{j,\sigma}:=(\mcl{R}_j)_\sigma$, with each map extended linearly to act on arbitrary operators.  Define
\begin{equation}\label{eq:qf-recovery-gap}
 \gamma_{\mcl{R}}(\sigma)
 :=\min_{\substack{X\in\mcl{B}(\hs),\ X=X^*\\ \Tr X=0,\ \norm{X}_2=1}}
 \sum_{j=1}^m\norm{X-\mcl{R}_{j,\sigma}(X)}_2^2.
\end{equation}
Then $\gamma_{\mcl{R}}(\sigma)>0$, and every $\rho\in\mcl{S}(\hs)$ satisfies
\begin{equation}\label{eq:qf-recovery-finite}
 \ent{\rho}{\sigma}
 \leq\frac{2}{\lambda_{\min}(\sigma)\gamma_{\mcl{R}}(\sigma)}
 \sum_{j=1}^m\entXA{\mcl{R}_j}{A_j}{\rho}{\sigma}.
\end{equation}
\end{proposition}
\begin{proof}
Let $\rho\in\mcl{S}(\hs)$ be fixed by every $\mcl{R}_{j,\sigma}$. Data processing for the partial trace and for recovery gives
\[
 \ent{\rho}{\sigma}
 =\ent{\rho_{A_j^c}}{\sigma_{A_j^c}},\qquad j=1,\ldots,m.
\]
Differentiating the modular identity in Theorem~\ref{thm:petz-sufficiency} at zero yields
\[
 \log\rho-\log\sigma
 =\identity_{A_j}\otimes
 (\log\rho_{A_j^c}-\log\sigma_{A_j^c}).
\]
Since the $A_j$ cover $\Lambda$, the intersection of these complementary algebras is $\mathds{C}\identity$. Hence $\log\rho-\log\sigma$ is scalar, and normalization gives $\rho=\sigma$.

If a nonzero traceless self-adjoint $X$ were fixed by all the maps, then $\sigma+\varepsilon X$ would be a distinct common fixed state for sufficiently small nonzero $\varepsilon$. Thus the minimum in \eqref{eq:qf-recovery-gap} is positive by compactness.

Now take arbitrary $\rho\in\mcl{S}(\hs)$. Since the maps fix $\sigma$, Pinsker's inequality and $\norm{\cdot}_1\geq\norm{\cdot}_2$ give
\begin{align*}
 \sum_{j=1}^m\entXA{\mcl{R}_j}{A_j}{\rho}{\sigma}
 &\geq\frac12\sum_{j=1}^m
 \norm{\rho-\mcl{R}_{j,\sigma}(\rho)}_2^2\\
 &\geq\frac{\gamma_{\mcl{R}}(\sigma)}2
 \norm{\rho-\sigma}_2^2.
\end{align*}
On the other hand, scalar Jensen's inequality in eigenbases of $\rho$ and $\sigma$, followed by $\log u\leq u-1$, gives
\begin{align*}
 \ent{\rho}{\sigma}
 &\leq\log\Tr[\rho^2\sigma^{-1}]
 \leq\Tr[\rho^2\sigma^{-1}]-1\\
 &=\Tr[(\rho-\sigma)^2\sigma^{-1}]
 \leq\lambda_{\min}(\sigma)^{-1}
       \norm{\rho-\sigma}_2^2.
\end{align*}
Combining the two estimates proves the claim.
\end{proof}

In particular, this applies to two complementary regions and to the single-site cover of a multipartite system. It needs no small-correlation assumption. The dependence on $\lambda_{\min}(\sigma)$ and $\gamma_{\mcl{R}}(\sigma)$, however, can make the estimate poor for large systems; it also gives no constant uniform in an auxiliary dimension.

\subsubsection{A bound from the interaction between recovery maps}

We next derive a quasi-factorization bound for two complementary regions, with a constant determined by the interaction between their recovery maps. Let $\mcl{R}_A$ and $\mcl{R}_B$ be recovery maps for $\identity_A\otimes\mcl{B}(\hs_B)$ and $\mcl{B}(\hs_A)\otimes\identity_B$, respectively, and write $\mcl{R}=(\mcl{R}_A,\mcl{R}_B)$. Recall the positive map
\[
 \mcl{T}_{\sigma_{AB}}(X)
 =\int_0^\infty(\sigma_{AB}+s\identity)^{-1}X
                   (\sigma_{AB}+s\identity)^{-1}\,ds.
\]
Define $\eta_{\mcl{R}}(\sigma_{AB})$ as the smallest nonnegative constant such that
\begin{equation}\label{eq:qf-recovery-interaction}
 \begin{split}
 \left|\Tr\left[
 \mcl{R}_{A,\sigma_{AB}}(\sigma_A\otimes Y_B)
 \mcl{T}_{\sigma_{AB}}\bigl(\mcl{R}_{B,\sigma_{AB}}(X_A\otimes\sigma_B)\bigr)
 \right]\right|\leq\eta_{\mcl{R}}(\sigma_{AB})\norm{X_A}_1\norm{Y_B}_1
 \end{split}
\end{equation}
for all traceless self-adjoint $X_A$ and $Y_B$. Here the recovery maps act on arbitrary operators by their linear extensions. This coefficient is finite in finite dimensions and depends on the reference state and the maps, but not on $\rho_{AB}$.

\begin{theorem}\label{thm:qf-recovery-axioms}
Let $\mcl{R}_A,\mcl{R}_B$ satisfy Definition~\ref{def:recovery-map} for the two complementary algebras above. For every $\rho_{AB},\sigma_{AB}\in\mcl{S}(\hs_A\otimes\hs_B)$,
\begin{equation}\label{eq:qf-recovery-axioms}
 \bigl(1-2\eta_{\mcl{R}}(\sigma_{AB})\bigr)\ent{\rho_{AB}}{\sigma_{AB}}
 \leq\entXA{\mcl{R}_A}{A}{\rho_{AB}}{\sigma_{AB}}
     +\entXA{\mcl{R}_B}{B}{\rho_{AB}}{\sigma_{AB}}.
\end{equation}
\end{theorem}

\begin{proof}
The expansion and normalization argument used in Theorem~\ref{thm:qfAB}, followed by Lieb's three-matrix trace inequality \cite[Theorem~7]{Lieb1973} and $\log x\leq x-1$, gives
\begin{align*}
 \ent{\rho_{AB}}{\sigma_{AB}}
   &-\entXA{\mcl{R}_A}{A}{\rho_{AB}}{\sigma_{AB}}
   -\entXA{\mcl{R}_B}{B}{\rho_{AB}}{\sigma_{AB}}\\
 &\quad\leq\log\Tr\exp\Bigl(\log\mcl{R}_{A,\sigma_{AB}}(\rho_{AB})
       +\log\mcl{R}_{B,\sigma_{AB}}(\rho_{AB})-\log\sigma_{AB}\Bigr)\\
 &\quad\leq\Tr\left[
 \mcl{R}_{A,\sigma_{AB}}(\rho_{AB})
 \mcl{T}_{\sigma_{AB}}\bigl(\mcl{R}_{B,\sigma_{AB}}(\rho_{AB})\bigr)
 \right]-1.
\end{align*}
By axioms~(\ref{ax:recovery-reference}) and~(\ref{ax:recovery-restriction}), both maps fix $\sigma_{AB}$, and their deviations depend only on the corresponding marginal:
\begin{align*}
 \mcl{R}_{A,\sigma_{AB}}(\rho_{AB})-\sigma_{AB}
 &=\mcl{R}_{A,\sigma_{AB}}\bigl(\sigma_A\otimes(\rho_B-\sigma_B)\bigr),\\
 \mcl{R}_{B,\sigma_{AB}}(\rho_{AB})-\sigma_{AB}
 &=\mcl{R}_{B,\sigma_{AB}}\bigl((\rho_A-\sigma_A)\otimes\sigma_B\bigr).
\end{align*}
These deviations have zero trace by axiom~(\ref{ax:recovery-cptp}). Using the identities
\[
 \mcl{T}_{\sigma_{AB}}(\sigma_{AB})=\identity,
 \qquad\Tr[\sigma_{AB}\mcl{T}_{\sigma_{AB}}(X)]=\Tr X,
\]
centering at $\sigma_{AB}$ gives
\begin{align*}
 &\Tr\left[
 \mcl{R}_{A,\sigma_{AB}}(\rho_{AB})
 \mcl{T}_{\sigma_{AB}}\bigl(\mcl{R}_{B,\sigma_{AB}}(\rho_{AB})\bigr)
 \right]-1\\
 &\qquad\qquad\quad=\Tr\Bigl[
 \mcl{R}_{A,\sigma_{AB}}\bigl(\sigma_A\otimes(\rho_B-\sigma_B)\bigr)\mcl{T}_{\sigma_{AB}}\Bigl(
 \mcl{R}_{B,\sigma_{AB}}\bigl((\rho_A-\sigma_A)\otimes\sigma_B\bigr)
 \Bigr)\Bigr]\\
 &\qquad\qquad\quad\leq\eta_{\mcl{R}}(\sigma_{AB})
       \norm{\rho_A-\sigma_A}_1\norm{\rho_B-\sigma_B}_1\\
 &\qquad\qquad\quad\leq2\eta_{\mcl{R}}(\sigma_{AB})\ent{\rho_{AB}}{\sigma_{AB}}.
\end{align*}
The last step uses Pinsker's inequality and data processing. Rearranging proves \eqref{eq:qf-recovery-axioms}.
\end{proof}

Thus $\eta_{\mcl{R}}(\sigma_{AB})<1/2$ gives the finite quasi-factorization constant
$\bigl(1-2\eta_{\mcl{R}}(\sigma_{AB})\bigr)^{-1}$.
At a product reference state, axiom~(\ref{ax:recovery-admissibility}) forces
$\mcl{R}_{A,\sigma_A\otimes\sigma_B}(\rho_{AB})=\sigma_A\otimes\rho_B$
and $\mcl{R}_{B,\sigma_A\otimes\sigma_B}(\rho_{AB})=\rho_A\otimes\sigma_B$.
Using
\[
 \mcl{T}_{\sigma_A\otimes\sigma_B}(X_A\otimes\sigma_B)
 =\mcl{T}_{\sigma_A}(X_A)\otimes\identity_B,
\]
the trace in \eqref{eq:qf-recovery-interaction} reduces to $\Tr X_A\,\Tr Y_B=0$.
Hence $\eta_{\mcl{R}}(\sigma_A\otimes\sigma_B)=0$, and exact tensorization follows for every admissible recovery family.

\subsubsection{An explicit coefficient for Petz recovery}

The following theorem is a particular case of Theorem~\ref{thm:qf-recovery-axioms}, obtained by taking the standard Petz recovery maps and bounding their interaction coefficient explicitly through modular powers of the reference state. We use the notation $D_X^P$ for $D_X^{\mcl{R}^P}$, and identify marginal operators with their embeddings on $AB$.

\begin{theorem}[{\cite[Theorem~7]{CLP18b}}]\label{thm:qf-petz-explicit}
Let $\rho_{AB},\sigma_{AB}\in\mcl{S}(\hs_A\otimes\hs_B)$. Define
\begin{align*}
 e_1(\sigma_{AB})
 &=\int_{\mathbb R}\beta(t)
 \norm{\sigma_B^{\frac{-1+it}{2}}\sigma_{AB}^{\frac{1-it}{2}}
             \sigma_A^{\frac{-1+it}{2}}-\identity_{AB}}_\infty
 \norm{\sigma_A^{-1/2}\sigma_{AB}^{\frac{1+it}{2}}\sigma_B^{-1/2}}_\infty\,dt,\\
 e_2(\sigma_{AB})
 &=\int_{\mathbb R}\beta(t)
 \norm{\sigma_B^{\frac{-1-it}{2}}\sigma_{AB}^{\frac{1+it}{2}}
             \sigma_A^{\frac{-1-it}{2}}-\identity_{AB}}_\infty\,dt,
 \qquad \beta(t)=\frac{\pi}{2(1+\cosh(\pi t))},\\
 \xi_P(\sigma_{AB})&=2\bigl(e_1(\sigma_{AB})+e_2(\sigma_{AB})\bigr).
\end{align*}
Then
\begin{equation}\label{eq:qf-petz-explicit}
 (1-\xi_P(\sigma_{AB}))\ent{\rho_{AB}}{\sigma_{AB}}
 \leq\entXA{P}{A}{\rho_{AB}}{\sigma_{AB}}
     +\entXA{P}{B}{\rho_{AB}}{\sigma_{AB}}.
\end{equation}
\end{theorem}

More precisely, the estimates in \cite[Section~4.3]{CLP18b}, using the modular integral representation of $\mcl{T}_{\sigma_{AB}}$ \cite[Lemma~3.4]{SBT17}, give
\[
 \eta_{(\mcl{R}_A^P,\mcl{R}_B^P)}(\sigma_{AB})
 \leq e_1(\sigma_{AB})+e_2(\sigma_{AB})
 =\frac{\xi_P(\sigma_{AB})}{2}.
\]
Substitution into Theorem~\ref{thm:qf-recovery-axioms} yields \eqref{eq:qf-petz-explicit}. The coefficient vanishes for $\sigma_{AB}=\sigma_A\otimes\sigma_B$; when $\xi_P(\sigma_{AB})<1$, it gives the quasi-factorization constant $(1-\xi_P(\sigma_{AB}))^{-1}$. Interchanging $A$ and $B$ gives another valid coefficient, and the smaller of the two can be used.

\subsubsection{Rotated and averaged recovery maps}

The rotated Petz maps and their averages satisfy Definition~\ref{def:recovery-map}. Therefore Proposition~\ref{prop:qf-recovery-finite} applies to them, including on multipartite covers. Theorem~\ref{thm:qf-recovery-axioms} also applies, with the interaction coefficient computed for the chosen pair of maps.

More generally, let $\mcl{R}_A^t,\mcl{R}_B^s$ be families of admissible recovery maps for the two complementary algebras and let $\mu,\nu$ be probability measures on $\mathbb R$. Define their averages at each fixed reference state $\sigma$ by
\[
 \overline{\mcl{R}}_{A,\sigma}=\int\mcl{R}_{A,\sigma}^t\,d\mu(t),\qquad
 \overline{\mcl{R}}_{B,\sigma}=\int\mcl{R}_{B,\sigma}^s\,d\nu(s).
\]
The resulting recovery families $\overline{\mcl{R}}_A,\overline{\mcl{R}}_B$ are admissible, and bilinearity of the trace in \eqref{eq:qf-recovery-interaction} gives
\begin{equation}\label{eq:qf-averaged-recovery}
 \eta_{(\overline{\mcl{R}}_A,\overline{\mcl{R}}_B)}(\sigma_{AB})
 \leq\int\!\!\int
 \eta_{(\mcl{R}_A^t,\mcl{R}_B^s)}(\sigma_{AB})\,d\mu(t)\,d\nu(s).
\end{equation}
If the right-hand side is less than $1/2$, it yields a quasi-factorization bound for the averaged maps. In particular, taking $\mcl{R}_A^t=\mcl{R}_A^{P,t}$ and $\mcl{R}_B^s=\mcl{R}_B^{P,s}$, with $d\mu(t)=\beta(t)\,dt$ and $d\nu(s)=\beta(s)\,ds$, gives $\overline{\mcl{R}}_A=\mcl{R}_A^{PT}$ and $\overline{\mcl{R}}_B=\mcl{R}_B^{PT}$. This observation follows from our interaction estimate; it is separate from the recovery-error bounds in \cite{junge_universal_2018,SBT17}.

For a multipartite product reference, let $\mcl{R}_{\{x\}}$ be an admissible recovery family for the complement algebra of each site $x\in\Lambda$. Axiom~(\ref{ax:recovery-admissibility}) gives $\mcl{R}_{\{x\},\sigma}(\rho)=\sigma_x\otimes\rho_{\Lambda\setminus\{x\}}$. Thus \eqref{eq:qf-product} gives the sharper exact bound
\[
 \ent{\rho}{\sigma}
 \leq\sum_{x\in\Lambda}\entXA{\mcl{R}_{\{x\}}}{\{x\}}{\rho}{\sigma},
 \qquad\sigma=\bigotimes_{x\in\Lambda}\sigma_x.
\]

\section{Weak quasi-factorization}
\label{sec:weak-quasi-factorization}

We distinguish a correction depending on $\rho$ from a uniform additive constant. The former gives weak quasi-factorization once it is bounded uniformly over the states under consideration. A nonzero uniform error also means that the estimate need not control arbitrarily small deviations from equilibrium; see the discussion in \cite{LaRacuente22}. In applications, one therefore seeks either a vanishing error or a family of estimates whose error can be reduced by changing the regions.

\subsection{Conditional relative entropy}
\label{subsec:weak-qf-expectations}

Let $\condexp_j=\condexp_\sigma^{\vna_j}$, $j=1,2$, and $\condexp_0=\condexp_\sigma^{\mcl{M}}$ be $\sigma$-preserving conditional expectations. Here we allow $\mcl{M}\subseteq\vna_1\cap\vna_2$, and hence $\condexp_j\condexp_0=\condexp_0\condexp_j=\condexp_0$.

\begin{definition}[Weak quasi-factorization]\label{def:ATcd}
The expectations satisfy $\operatorname{AT}(c,d)$, with $c\geq1$ and $d\geq0$, if, for every $\rho\in\mcl{S}(\hs)$,
\begin{equation}\label{eq:weak-qf-definition}
 \ent{\rho}{\condexp_0^*(\rho)}
 \leq c\left(\ent{\rho}{\condexp_1^*(\rho)}
             +\ent{\rho}{\condexp_2^*(\rho)}\right)+d.
\end{equation}
Equivalently, these are the conditional relative entropies $D_{\mcl{B}(\hs)|\mcl{M}}$, $D_{\mcl{B}(\hs)|\vna_1}$ and $D_{\mcl{B}(\hs)|\vna_2}$. The case $d=0$ is the quasi-factorization considered previously, also called \emph{strong} approximate tensorization. For a finite family, the two terms on the right are replaced by the sum over that family.
\end{definition}

\subsubsection{A trace correction and a uniform max-relative-entropy bound}

Allowing an additive correction extends the exact factorization bound for commuting squares to general pairs of conditional expectations. For that, we first keep the logarithmic remainder in the three-matrix argument. The following two statements are \cite[Lemma~1 and Corollary~1]{BCR22}.

\begin{lemma}\label{lem:weak-qf-expectations}
For every $\rho\in\mcl{S}(\hs)$,
\begin{equation}\label{eq:weak-qf-expectations}
 \begin{split}
 \ent{\rho}{\condexp_0^*(\rho)}
 &\leq\ent{\rho}{\condexp_1^*(\rho)}
       +\ent{\rho}{\condexp_2^*(\rho)} +\log\Tr\left[\condexp_1^*(\rho)
 \mcl{T}_{\condexp_0^*(\rho)}\bigl(\condexp_2^*(\rho)\bigr)\right],
 \end{split}
\end{equation}
where $\mcl{T}_\omega(X)=\int_0^\infty(\omega+s\identity)^{-1}X(\omega+s\identity)^{-1}\,ds$.
\end{lemma}
\begin{proof}
Let $\rho\in\mcl{S}(\hs)$. Expanding the difference on the left gives
\[
 \begin{split}
 \ent{\rho}{\condexp_0^*(\rho)}
 -\ent{\rho}{\condexp_1^*(\rho)}-&\ent{\rho}{\condexp_2^*(\rho)}\\
 &\quad=\Tr\left[\rho\left(\log\condexp_1^*(\rho)
 +\log\condexp_2^*(\rho)-\log\condexp_0^*(\rho)-\log\rho\right)\right]\\
 &\quad\leq\log\Tr\left[\exp\left(\log\condexp_1^*(\rho)
 +\log\condexp_2^*(\rho)-\log\condexp_0^*(\rho)\right)\right].
 \end{split}
\]
The last inequality follows by normalizing the exponential and using nonnegativity of relative entropy. Lieb's three-matrix inequality \cite{Lieb1973} gives the claimed bound.
\end{proof}

The next corollary turns this state-dependent trace correction into a uniform additive constant by comparing the output of $\condexp_1^*\condexp_2^*$ with states fixed by $\condexp_0^*$ in max-relative entropy.

\begin{corollary}\label{cor:weak-qf-expectations}
With $D_{\max}(\omega\Vert\eta)=\inf\{\log a:\omega\leq a\eta\}$, set
\[
 d=\sup_{\substack{\omega\in\mcl{S}(\hs)\\\condexp_2^*(\omega)=\omega}}
   \inf_{\substack{\eta\in\mcl{S}(\hs)\\\condexp_0^*(\eta)=\eta}}
   D_{\max}\bigl(\condexp_1^*(\omega)\Vert\eta\bigr).
\]
Then $\operatorname{AT}(1,d)$ holds.
\end{corollary}

Interchanging the two expectations gives another admissible constant, so one can take the smaller of the two. Choosing $\eta=\sigma$ gives the elementary bound $d\leq\log\lambda_{\min}(\sigma)^{-1}$. Under the commuting-square condition $\condexp_1\condexp_2=\condexp_2\condexp_1=\condexp_0$, the state $\condexp_1^*(\omega)$ is already fixed by $\condexp_0^*$, and therefore $d=0$.

\subsubsection{Tracial expectations and overlap of subalgebras}
\label{subsubsec:weak-qf-algebraic}

For tracial expectations, an algebraic entropy inequality gives a uniform defect directly in terms of the two ranges. As before, the superscript $(0)$ denotes the tracial expectation onto the indicated algebra.

\begin{proposition}[{\cite[Corollary~3.4]{GaoJungeLaRacuente23}}]\label{prop:weak-qf-algebraic}
Define
\[
 \chi=\sup\left\{\Tr[ab]:\ a\in(\vna_1)_+,\ b\in(\vna_2)_+,\
 \condexp_0^{(0)}(a)=\identity,\ \Tr b=1\right\}.
\]
Then the tracial expectations satisfy $\operatorname{AT}(1,\log\chi)$.
\end{proposition}
\begin{proof}
The cited result states that
\[
 S\bigl(\condexp_1^{(0)*}(\rho)\bigr)
 +S\bigl(\condexp_2^{(0)*}(\rho)\bigr)
 \geq S(\rho)+S\bigl(\condexp_0^{(0)*}(\rho)\bigr)-\log\chi.
\]
For tracial expectations,
$\ent{\rho}{\condexp_j^{(0)*}(\rho)}=S(\condexp_j^{(0)*}(\rho))-S(\rho)$,
which proves the assertion.
\end{proof}

For example, let $\vna_1,\vna_2$ be the diagonal algebras in orthonormal bases $(e_i)$ and $(f_j)$ of an $n$-dimensional space, and take $\mcl{M}=\mathds{C}\identity$. Then
\[
 \chi=n\max_{i,j}|\langle e_i,f_j\rangle|^2,
 \qquad d=\log\left(n\max_{i,j}|\langle e_i,f_j\rangle|^2\right).
\]
Thus mutually unbiased bases give zero defect. This connects weak quasi-factorization with entropic uncertainty relations; multiplicative estimates for related dephasing maps are discussed in \cite[Section~IV.A]{LaRacuente22}.

\subsubsection{Changing the reference state}

To apply the tracial estimates to a nontracial reference state, we need to control how both the multiplicative constant and the additive defect change. The change-of-measure argument used above provides such a comparison.

\begin{proposition}[{\cite[Proposition~2]{BCR22}}]\label{prop:weak-qf-change-measure}
If the tracial expectations onto $\mcl{M},\vna_1,\vna_2$ satisfy $\operatorname{AT}(1,d_0)$, then the $\sigma$-preserving expectations satisfy $\operatorname{AT}(c,d)$ with
\[
 c=\frac{\lambda_{\max}(\sigma)}{\lambda_{\min}(\sigma)},
 \qquad d=c\,d_0.
\]
\end{proposition}

\begin{proof}[Proof outline]
We use the change-of-measure comparison of Proposition~\ref{prop:qf-change-measure}, following \cite[Appendix~B.1]{BCR22}. For a state $\rho$, put $X=\sigma^{-1/2}\rho\sigma^{-1/2}$. Applying the tracial inequality to $X/\Tr X$ and multiplying by $\Tr X$ introduces the additive term $d_0\Tr X$. The entropy comparison therefore gives
\[
 \ent{\rho}{\condexp_0^*(\rho)}
 \leq c\sum_{j=1}^2\ent{\rho}{\condexp_j^*(\rho)}
 +\lambda_{\max}(\sigma)d_0\Tr(\sigma^{-1}\rho).
\]
Since $\Tr(\sigma^{-1}\rho)\leq\lambda_{\min}(\sigma)^{-1}$, the last term is at most $c\,d_0$. The normalization factor $\Tr X$ gives the additive constant $c\,d_0$ in place of $(\dim\hs)\lambda_{\max}(\sigma)d_0$ as originally appeared in \cite[Proposition~2]{BCR22}.

\end{proof}

When $d_0=0$, this recovers Proposition~\ref{prop:qf-change-measure} with initial multiplicative constant one. For $d_0>0$, both constants must be tracked when the dimension grows.

\subsubsection{A pinching remainder}

For each fixed $\rho$, the pinching used below is a tracial conditional expectation, generalizing the dephasing maps considered in Subsection~\ref{subsubsec:weak-qf-algebraic}, with projections chosen according to the reduced block states of $\rho$. The additive remainder in the following estimate measures the information lost under this map. To define it, write
\[
 \hs=\bigoplus_r\hs_r\otimes\mcl{K}_r,
 \qquad\mcl{M}=\bigoplus_r\mcl{B}(\hs_r)\otimes\identity_{\mcl{K}_r},
 \qquad\condexp_0^*(\rho)=\sum_r\Tr_{\mcl{K}_r}[P_r\rho P_r]\otimes\omega_r,
\]
where $P_r$ is the block projection and $\omega_r\in\mcl{S}(\mcl{K}_r)$. On each diagonal block,
$\condexp_j=\operatorname{id}_{\mcl{B}(\hs_r)}\otimes\condexp_{j,r}$ for $j=1,2$,
and $\condexp_{0,r}(X)=\Tr(\omega_rX)\identity$. Choose an orthonormal eigenbasis $(e_{r,k})_k$ of $\Tr_{\mcl{K}_r}[P_r\rho P_r]$ and define
\[
 Q_{r,k}=|e_{r,k}\rangle\langle e_{r,k}|\otimes\identity_{\mcl{K}_r},
 \qquad\mcl{P}_\rho(X)=\sum_{r,k}Q_{r,k}XQ_{r,k}.
\]
The projectors here are embedded in their respective blocks of $\hs$.

\begin{theorem}[{\cite[Theorem~2, Eq.~(3.10)]{BCR22}}]\label{thm:weak-qf-pinching}
Suppose that
\[
 \theta=\max_r\norm{\condexp_{1,r}\condexp_{2,r}-\condexp_{0,r}:
 L_1(\omega_r)\longrightarrow L_\infty}<1,
 \qquad\norm{X}_{1,\omega_r}=\norm{\omega_r^{1/2}X\omega_r^{1/2}}_1.
\]
Then
\begin{equation}\label{eq:weak-qf-pinching}
 \ent{\rho}{\condexp_0^*(\rho)}
 \leq\frac{\ent{\rho}{\condexp_1^*(\rho)}+\ent{\rho}{\condexp_2^*(\rho)}}{1-\theta}
 +\ent{\rho}{\mcl{P}_\rho(\rho)}.
\end{equation}
\end{theorem}

The pinching inequality $\rho\leq m\mcl{P}_\rho(\rho)$, with $m=\sum_r\dim\hs_r$, implies
\[
 \ent{\rho}{\mcl{P}_\rho(\rho)}\leq\log m\leq\log\dim\hs.
\]
Consequently, \eqref{eq:weak-qf-pinching} yields a uniform additive constant, while retaining the remainder gives zero error on states fixed by $\mcl{P}_\rho$. This distinction is useful in the peeling arguments of \cite{capel_modified_2021}: one first applies suitable local expectations and then proves factorization for the resulting states. See also \cite[Theorem~5.9]{KACR25} for such a restricted-state tensorization in commuting lattice systems.

The spectral and completely positive order criteria of \cite{GaoRouze22,LaRacuente22,GJLL23}, recalled in the previous section, instead give estimates with no additive remainder. Thus the pinching bound is most useful when it offers better quantitative control or when its remainder can be eliminated using the structure of the states.

\subsubsection{Davies expectations and decay of matrix-valued conditional mutual information}

Let $\Lambda=\{-L,\ldots,L\}^2\subset\mathbb Z^2$ be a finite two-dimensional lattice, with Hilbert space $\hs=\hs_\Lambda=\bigotimes_{x\in\Lambda}\hs_x$, and let $\sigma$ be the Gibbs state of a commuting Hamiltonian of interaction range $r$.

Throughout this subsection, we use the dissipative Davies generators of \cite[Appendix~A.3]{capel_quasi-optimal_2025}. At each site $x$, the self-adjoint coupling operators form a Hilbert--Schmidt orthonormal basis of the full algebra $\mcl{B}(\hs_x)$ and are embedded into the full system by tensoring with identities. The rates at every occurring Bohr frequency (energy difference) are strictly positive and satisfy thermal detailed balance. For $X\subseteq\Lambda$, let $\mcl{L}_X$ be the sum of these single-site generators over $x\in X$, acting on observables, and write $\condexp_X=\lim_{t\to\infty}e^{t\mcl{L}_X}$; its action on states is $\condexp_X^*$. These coupling assumptions ensure that the comparison \eqref{eq:qf-davies-block-comparison} applies to every region $X$ \cite[Theorem~1 and Proposition~7]{BCR22}.

The lattice geometry now plays a direct role in the factorization bound: it allows us to construct a cover by blocks with controlled sizes, separations and overlaps. The overlap widths control the distances between the regions entering each additive defect. Under decay of matrix-valued conditional mutual information with distance, increasing these widths therefore gives a smaller bound on the error.

For disjoint regions $A,C,D\subseteq\Lambda$, define the \emph{matrix-valued conditional mutual information} by
\begin{equation}\label{eq:mcmi-definition}
 H_\sigma(A:C\mid D)
 =\log\sigma_{ACD}+\log\sigma_D-\log\sigma_{AD}-\log\sigma_{CD},
\end{equation}
with identities on the missing tensor factors understood \cite[Definition~2.3]{capel_quasi-optimal_2025}. Its expectation is the usual conditional mutual information:
\[
 \Tr\left[\sigma_{ACD}H_\sigma(A:C\mid D)\right]
 =I_\sigma(A:C\mid D)\leq\norm{H_\sigma(A:C\mid D)}_\infty.
\]
The operator itself need not be positive. Controlling its norm therefore controls more than this average.

\begin{figure}[htbp]
 \centering
 \begin{tikzpicture}[x=.6cm,y=.6cm]
  \fill[figConditioning] (0,0) rectangle (9,9);
  \fill[figRegionA] (1,1) rectangle (3,8);
  \fill[figRegionB] (3,1) rectangle (6,8);
  \fill[figRegionC] (6,1) rectangle (8,8);
  \foreach \u in {.5,1.5,...,8.5} {
   \draw[black!18,very thin] (.5,\u) -- (8.5,\u);
   \draw[black!18,very thin] (\u,.5) -- (\u,8.5);
   \foreach \v in {.5,1.5,...,8.5} {
    \fill[black!35] (\u,\v) circle (.6pt);
   }
  }
  \draw (0,0) rectangle (9,9);
  \draw (1,1) rectangle (8,8);
  \draw (3,1) -- (3,8);
  \draw (6,1) -- (6,8);
  \node[fill=figRegionA,inner sep=3pt] at (2,5) {$A$};
  \node[fill=figRegionB,inner sep=3pt] at (4.5,5) {$B$};
  \node[fill=figRegionC,inner sep=3pt] at (7,5) {$C$};
  \node[fill=figConditioning,inner sep=2pt] at (4.5,8.5) {$D$};
  \node[anchor=south east] at (9,9.1) {$\Lambda$};
  \draw[<->,thick] (2.5,2.5) -- (6.5,2.5);
  \node[fill=figRegionB,inner sep=2pt,font=\small] at (4.5,3.1)
   {$\operatorname{dist}(A,C)$};
 \end{tikzpicture}
 \caption{One possible geometry for $\Lambda=A\sqcup B\sqcup C\sqcup D$ on a two-dimensional lattice. The buffer $B$ separates $A$ and $C$ and is traced out, while the surrounding region $D$ is retained as the conditioning region in $H_\sigma(A:C\mid D)$.}
 \label{fig:mcmi-geometry}
\end{figure}
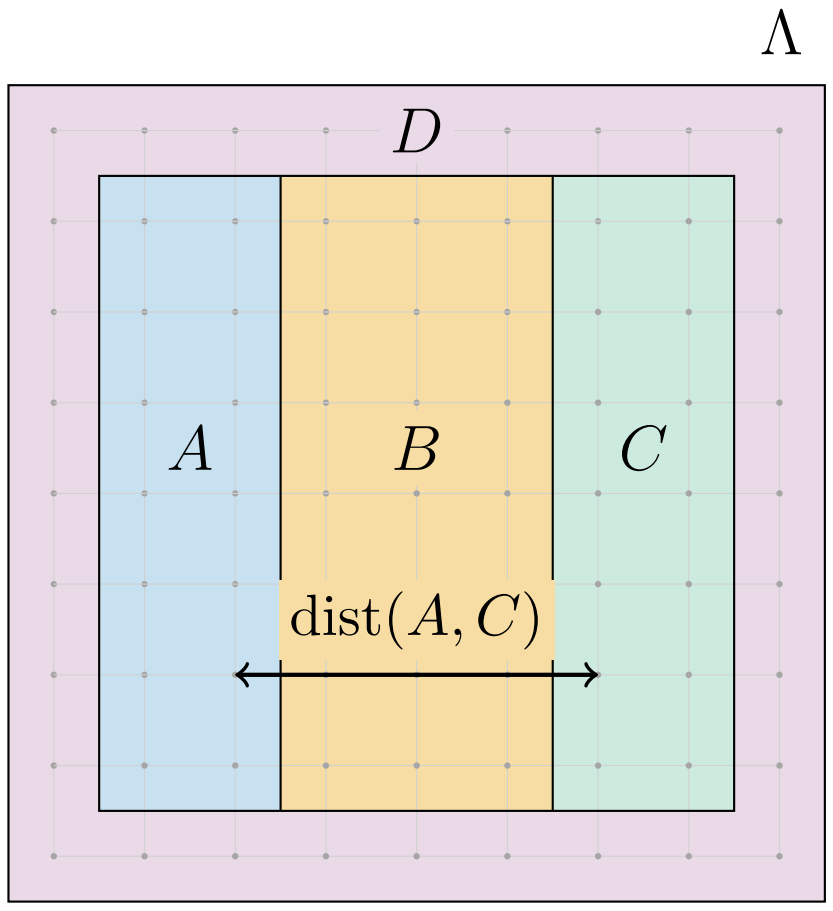

We say that $\sigma$ has exponential MCMI decay with constants $K,\xi>0$ if, for every partition $\Lambda=A\sqcup B\sqcup C\sqcup D$,
\begin{equation}\label{eq:mcmi-decay}
 \norm{H_\sigma(A:C\mid D)}_\infty
 \leq K|\Lambda|e^{-\operatorname{dist}(A,C)/\xi}.
\end{equation}
Here $B$ is traced out and $D$ is the conditioning region, as illustrated in Figure~\ref{fig:mcmi-geometry}. This assumption is established, for example, at sufficiently high temperature for marginal commuting Hamiltonians in \cite[Theorem~2.4 and Appendix~D.2]{capel_quasi-optimal_2025}. In this class, the algebra generated by the interaction terms is commutative and closed under partial traces, with the resulting operators embedded back into the full algebra.

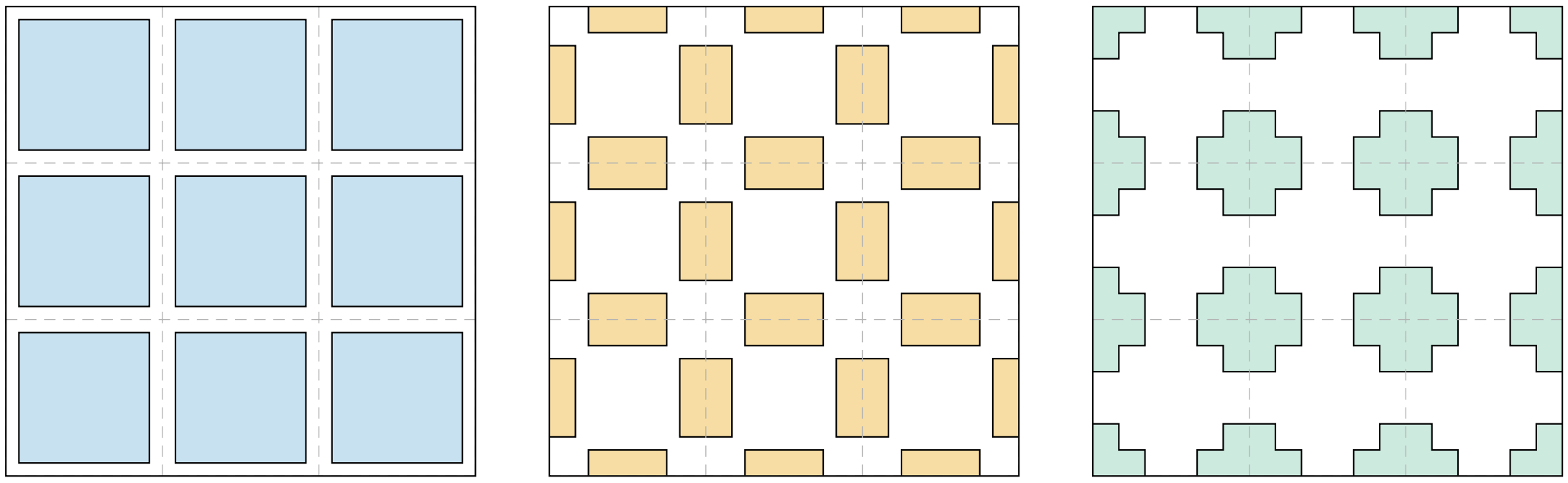
\begin{figure}[htbp]
 \centering
 \begin{tikzpicture}[x=.24cm,y=.24cm,line width=.4pt,font=\small]
  \begin{scope}
   \begin{scope}
    \clip (0,0) rectangle (18,18);
    \foreach \x in {0,6,12} {
     \foreach \y in {0,6,12} {
      \draw[fill=figRegionA] (\x+.5,\y+.5) rectangle (\x+5.5,\y+5.5);
     }
    }
   \end{scope}
   \draw[gray!60,densely dashed,line width=.25pt,step=6] (0,0) grid (18,18);
   \draw (0,0) rectangle (18,18);
   \node at (9,-1.5) {(a) Squares $S_i$};
  \end{scope}
  \begin{scope}[xshift=5cm]
   \begin{scope}
    \clip (0,0) rectangle (18,18);
    \foreach \x in {0,6,12} {
     \foreach \y in {0,6,12,18} {
      \draw[fill=figRegionB] (\x+1.5,\y-1) rectangle (\x+4.5,\y+1);
      \draw[fill=figRegionB] (\y-1,\x+1.5) rectangle (\y+1,\x+4.5);
     }
    }
   \end{scope}
   \draw[gray!60,densely dashed,line width=.25pt,step=6] (0,0) grid (18,18);
   \draw (0,0) rectangle (18,18);
   \node at (9,-1.5) {(b) Edge strips $E_j$};
  \end{scope}
  \begin{scope}[xshift=10cm]
   \begin{scope}
    \clip (0,0) rectangle (18,18);
    \foreach \x in {0,6,12,18} {
     \foreach \y in {0,6,12,18} {
      \draw[fill=figRegionC]
       (\x-2,\y-1) -- (\x-1,\y-1) -- (\x-1,\y-2)
       -- (\x+1,\y-2) -- (\x+1,\y-1) -- (\x+2,\y-1)
       -- (\x+2,\y+1) -- (\x+1,\y+1) -- (\x+1,\y+2)
       -- (\x-1,\y+2) -- (\x-1,\y+1) -- (\x-2,\y+1) -- cycle;
     }
    }
   \end{scope}
   \draw[gray!60,densely dashed,line width=.25pt,step=6] (0,0) grid (18,18);
   \draw (0,0) rectangle (18,18);
   \node at (9,-1.5) {(c) Vertex patches $V_k$};
  \end{scope}
 \end{tikzpicture}
 \caption{A schematic two-dimensional cover by three families of regions, shown separately on the same coarse grid. The squares $S_i$, edge strips $E_j$ and vertex patches $V_k$ together cover the lattice domain. Regions within each family are separated, while regions from different families overlap. The outer boundary is aligned with the coarse grid. Boundary shapes are schematic; the theorem uses the boundary blocks of the compatible-box construction cited in the text}
 \label{fig:geometric-cover-2d}
\end{figure}

Write $\condexp_X$ for the limiting Davies expectation on observables associated with updates in $X\subseteq\Lambda$; its action on states is $\condexp_X^*$. Fix an overlap width $b\geq r$ and a cell size $\ell>4(r+b)$, chosen so that the cover fits inside $\Lambda$. The construction of \cite[Lemma~2.7 and Appendix~B.2]{capel_quasi-optimal_2025} gives three families of blocks: squares $S_i$ in the cell interiors, strips $E_j$ along the cell edges, and small patches $V_k$ around the vertices, as shown in Figure~\ref{fig:geometric-cover-2d}. We write $\mcl{C}$ for the collection of all these blocks.

The three families together cover $\Lambda$. Blocks within each family are separated by more than $r$, and every block contains at most $\ell^2$ sites. The overlaps between families have width at least $b$: they allow us to split off the square interiors first, and then the interiors of the edge strips, leaving the vertex patches. In each split, the overlap separates the two nonoverlapping parts by at least $b$.

The proof of the following theorem uses the marginal-difference inequality in Lemma~\ref{lem:weak-entropy-factorization}, proved in Subsection~\ref{subsec:weak-qf-marginal}. We present the conditional-expectation result first to follow the same order as in the rest of the chapter.

\begin{theorem}[{\cite[Theorem~B.3]{capel_quasi-optimal_2025}}]\label{thm:wAT-Davies}
For the Davies expectations with the coupling and rate assumptions specified above, assume the MCMI decay condition \eqref{eq:mcmi-decay}. For the two-dimensional cover $\mcl{C}$ obtained from a compatible tiling as above, and every $\rho\in\mcl{S}(\hs)$,
\begin{equation}\label{eq:weak-qf-Davies}
 \ent{\rho}{\sigma}
 \leq\sum_{X\in\mcl{C}}\ent{\rho}{\condexp_X^*(\rho)}
 +2K|\Lambda|e^{-b/\xi}.
\end{equation}
\end{theorem}

\begin{proof}[Proof outline]
Let $S,E,V$ denote the unions of the square, strip and vertex blocks, respectively. Apply Lemma~\ref{lem:weak-entropy-factorization} below first to $\Lambda=S\cup(E\cup V)$ and then to $E\cup V$. Each split has an overlap of width at least $b$, so \eqref{eq:mcmi-decay} gives
\begin{align*}
 \ent{\rho}{\sigma}
 &\leq\entdA{S}{\rho}{\sigma}+\entdA{E\cup V}{\rho}{\sigma}
       +K|\Lambda|e^{-b/\xi},\\
 \entdA{E\cup V}{\rho}{\sigma}
 &\leq\entdA{E}{\rho}{\sigma}+\entdA{V}{\rho}{\sigma}
       +K|\Lambda|e^{-b/\xi}.
\end{align*}
In the second split, $\Lambda\setminus(E\cup V)$ is retained as the conditioning region. The comparison \eqref{eq:qf-davies-block-comparison} bounds each of $D_S^d,D_E^d,D_V^d$ by the relative entropy to the corresponding Davies expectation. Within each family, the separated blocks give commuting Davies expectations, so Proposition~\ref{prop:qf-commuting-square} splits these three terms into the sum over $\mcl{C}$.
\end{proof}

For a prescribed error $\varepsilon>0$, taking
\[
 b\geq\max\left\{r,\xi\log\frac{2K|\Lambda|}{\varepsilon}\right\}
\]
gives additive defect at most $\varepsilon$, provided the covering parameters fit inside the lattice. Thus a logarithmic increase in overlap yields a chosen error tolerance. The terms in \eqref{eq:weak-qf-Davies} are the usual conditional relative entropies associated with the fixed-point algebras of the local Davies dynamics.

The construction extends to $n$ dimensions with $n+1$ families, associated with cells of dimensions $n,n-1,\ldots,0$. Choosing $\ell>2n(r+b)$, the same argument uses $n$ splitting steps and gives an additive defect at most $nK|\Lambda|e^{-b/\xi}$.

These quasi-factorizations turn local entropy-dissipation estimates into global convergence bounds. The approach based on weak quasi-factorization was already developed in \cite{BCR22,capel_modified_2021}. Related factorization arguments have yielded modified logarithmic Sobolev inequalities (MLSIs) for local GNS-symmetric dynamics of commuting lattice Hamiltonians \cite{capel_modified_2021,BCGLPR24,KACR25}; retaining the defect in the construction above gives weak MLSIs \cite{capel_quasi-optimal_2025}. This line of work predates \cite{ChenDingLu26}, who obtain size-independent MLSIs for a Gaussian KMS Gibbs sampler, including noncommuting models, by a different method.

\subsection{Conditional relative entropy by marginal difference}
\label{subsec:weak-qf-marginal}

For marginal differences, the additive correction can be expressed directly through logarithms of the reference state and its marginals, and hence these estimates do not require a conditional expectation onto each complement algebra. We begin with the splitting inequality used in the lattice argument above.

\subsubsection{Overlapping regions and a logarithmic defect}

\begin{lemma}[{\cite[Lemma~2.6 and Appendix~B.1]{capel_quasi-optimal_2025}}]\label{lem:weak-entropy-factorization}
Let $\hs=\hs_A\otimes\hs_B\otimes\hs_C\otimes\hs_D$ and $\rho,\sigma\in\mcl{S}(\hs)$. Then
\begin{equation}\label{eq:weak-entropy-factorization}
 \entdA{ABC}{\rho}{\sigma}
 \leq\entdA{AB}{\rho}{\sigma}+\entdA{BC}{\rho}{\sigma}
 +\norm{H_\sigma(A:C\mid D)}_\infty.
\end{equation}
\end{lemma}
\begin{proof}
Let $\rho\in\mcl{S}(\hs)$. Expanding the conditional relative entropies and applying data processing under $\Tr_B$ gives
\begin{align*}
 \entdA{ABC}{\rho}{\sigma}&-\entdA{AB}{\rho}{\sigma}-\entdA{BC}{\rho}{\sigma}\\
 &\quad=\ent{\rho_{AD}}{\sigma_{AD}}+\ent{\rho_{CD}}{\sigma_{CD}}
       -\ent{\rho_D}{\sigma_D}-\ent{\rho}{\sigma}\\
 &\quad\leq\ent{\rho_{AD}}{\sigma_{AD}}+\ent{\rho_{CD}}{\sigma_{CD}}
       -\ent{\rho_D}{\sigma_D}-\ent{\rho_{ACD}}{\sigma_{ACD}}\\
 &\quad=-I_\rho(A:C\mid D)+\Tr\left[\rho_{ACD}H_\sigma(A:C\mid D)\right]\\
 &\quad\leq\norm{H_\sigma(A:C\mid D)}_\infty.
\end{align*}
The last step uses strong subadditivity and trace-norm/operator-norm duality.
\end{proof}

If $B$ is empty, the data-processing step is an equality. In particular, $H_\sigma(A:C\mid D)=0$ recovers exact conditional factorization for a quantum Markov state. With both $B$ and $D$ empty, and relabelling $C$ as $B$, we obtain the bipartite bound
\begin{equation}\label{eq:weak-qf-logarithmic}
 \ent{\rho_{AB}}{\sigma_{AB}}
 \leq\entdA{A}{\rho_{AB}}{\sigma_{AB}}
 +\entdA{B}{\rho_{AB}}{\sigma_{AB}}
 +\norm{\log\sigma_{AB}-\log(\sigma_A\otimes\sigma_B)}_\infty.
\end{equation}
Its defect is independent of $\rho$ and vanishes at a product reference.

The overlapping version can be iterated: if $V_j=A_j\cup B_j$ and the next splitting is applied to $B_j$, each step adds its own MCMI norm. This yields multipartite estimates with the sum of the local defects.

For local Davies expectations, applying \eqref{eq:cre-comparison-enlarged-region} term by term transfers these bounds to conditional-expectation entropies, and transfers bounds of the latter form to marginal differences on enlarged regions. Both the multiplicative constant and the additive error are preserved.

\subsubsection{A continuity bound near product states}

An alternative estimate uses trace distances from product marginals. The following is \cite[Corollary~7.6]{BCGP23}, as stated in the corrected arXiv version, and follows from the relative-entropy continuity bound in their Theorem~5.13.

\begin{proposition}\label{cor:weak-qf-relative-entropy}
For $\rho_{AB},\sigma_{AB}\in\mcl{S}(\hs_A\otimes\hs_B)$, denote
\begin{equation*}
     \varepsilon=\frac12\norm{\rho_{AB}-\rho_A\otimes\rho_B}_1,
 \qquad\delta=\frac12\norm{\sigma_{AB}-\sigma_A\otimes\sigma_B}_1.
\end{equation*}
Then, we have
\[
 \ent{\rho_{AB}}{\sigma_{AB}}
 \leq\entdA{A}{\rho_{AB}}{\sigma_{AB}}
 +\entdA{B}{\rho_{AB}}{\sigma_{AB}}
 +\xi_{\mathrm{RE}}(\rho_{AB},\sigma_{AB}),
\]
where
\begin{align*}
 \xi_{\mathrm{RE}}
 &=\left(\sqrt2-\log\widetilde m\right)\sqrt\varepsilon
 +3\frac{\log\widetilde m^{-1}}{1-\widetilde m}\delta
 +2\log\left(1+\frac{\delta}{1-\widetilde m+\delta}\frac1{\widetilde m}\right),\\
 \widetilde m
 &=\frac12\min\left\{\lambda_{\min}(\sigma_A)\lambda_{\min}(\sigma_B),
                         \lambda_{\min}(\sigma_{AB})\right\}.\end{align*}
\end{proposition}
\begin{proof}[Proof outline]
The assertion is equivalent to
\[
 \ent{\rho_A}{\sigma_A}+\ent{\rho_B}{\sigma_B}
 \leq\ent{\rho_{AB}}{\sigma_{AB}}+\xi_{\mathrm{RE}}.
\]
The left-hand side equals $\ent{\rho_A\otimes\rho_B}{\sigma_A\otimes\sigma_B}$. Applying the continuity estimate of \cite[Theorem~5.13]{BCGP23} to this product pair and $(\rho_{AB},\sigma_{AB})$ gives the stated correction.
\end{proof}

This correction is useful when both states are close to products. A uniform defect follows by using $\varepsilon\leq1$, although that loses the improvement for nearly product inputs. Unlike \eqref{eq:weak-qf-logarithmic}, this continuity estimate need not have zero correction when only $\sigma_{AB}$ is a product.

\subsubsection{A channel-based relative-entropy superadditivity bound}

\cite[Theorem~5.2]{GaoJungeLaRacuente23} proves a corrected strong-subadditivity inequality for relative entropy under two quantum channels. Here we particularise it to partial traces, which gives another weak quasi-factoriza\-tion by marginal differences.

\begin{proposition}\label{prop:weak-qf-channel}
Let $\rho_{AB},\sigma_{AB}\in\mcl{S}(\hs_A\otimes\hs_B)$. Then
\begin{equation}\label{eq:weak-qf-channel}
 \begin{split}
 \ent{\rho_{AB}}{\sigma_{AB}}
 \leq\entdA{A}{\rho_{AB}}{\sigma_{AB}}
 +\entdA{B}{\rho_{AB}}{\sigma_{AB}}+\kappa(\rho_A,\sigma_{AB}),
 \end{split}
\end{equation}
where
\[
 \kappa(\rho_A,\sigma_{AB})
 =\int_{\mathbb R}\frac{\pi}{2(\cosh(\pi t)+1)}\log c_t\,dt,
\]
for
\[
 c_t=\norm{\sigma_B^{-1/2}\Tr_A\left[
 \left(\rho_A^{(1+it)/2}\sigma_A^{-(1+it)/2}\otimes\identity_B\right)
 \sigma_{AB}
 \left(\sigma_A^{-(1-it)/2}\rho_A^{(1-it)/2}\otimes\identity_B\right)\right]
                  \sigma_B^{-1/2}}_\infty.
\]
\end{proposition}
\begin{proof}
For the proof, abbreviate $K_t=\rho_A^{(1+it)/2}\sigma_A^{-(1+it)/2}\otimes\identity_B$.
Apply \cite[Theorem~5.2]{GaoJungeLaRacuente23} with channels $\Tr_B$ and $\Tr_A$, and with the scalar output algebra on $B$. Its $\sigma_B$-preserving expectation is $b\mapsto\Tr(\sigma_B b)\identity_B$. The correction in that theorem is the integral above because
\[
 \sup_{b\geq0,\,\Tr(\sigma_B b)=1}
 \Tr\left[(\identity_A\otimes b)K_t\sigma_{AB}K_t^*\right]=c_t.
\]
The resulting inequality is
\[
 \ent{\rho_A}{\sigma_A}+\ent{\rho_B}{\sigma_B}
 \leq\ent{\rho_{AB}}{\sigma_{AB}}+\kappa(\rho_A,\sigma_{AB}),
\]
which is equivalent to \eqref{eq:weak-qf-channel}.
\end{proof}

At a product reference state, $\Tr_A[K_t(\sigma_A\otimes\sigma_B)K_t^*]=\sigma_B$, so the correction vanishes. For a general reference state, the same identity gives the following uniform consequence.

\begin{corollary}\label{cor:weak-qf-max-product}
For every $\rho_{AB},\sigma_{AB}\in\mcl{S}(\hs_A\otimes\hs_B)$,
\begin{equation}\label{eq:weak-qf-max-product}
 \begin{split}
 \ent{\rho_{AB}}{\sigma_{AB}}
 &\leq\entdA{A}{\rho_{AB}}{\sigma_{AB}}
 +\entdA{B}{\rho_{AB}}{\sigma_{AB}}+D_{\max}(\sigma_{AB}\Vert\sigma_A\otimes\sigma_B).
 \end{split}
\end{equation}
\end{corollary}
\begin{proof}
Set $\alpha=\exp D_{\max}(\sigma_{AB}\Vert\sigma_A\otimes\sigma_B)$. Since $\sigma_{AB}\leq\alpha\sigma_A\otimes\sigma_B$, positivity gives
\[
 \Tr_A[K_t\sigma_{AB}K_t^*]
 \leq\alpha\Tr_A[K_t(\sigma_A\otimes\sigma_B)K_t^*]
 =\alpha\sigma_B.
\]
Thus $c_t\leq\alpha$ for every $t$, and $\kappa\leq\log\alpha$. Substituting in \eqref{eq:weak-qf-channel} proves the bound.
\end{proof}

This corollary can also be obtained directly from the entropy expansion in Lemma~\ref{lem:weak-entropy-factorization}, using operator monotonicity of the logarithm. It is a uniform consequence of that argument, but $\kappa(\rho_A,\sigma_{AB})$ may give a smaller correction for a particular input.

\subsubsection{Related results for the Belavkin--Staszewski relative entropy}

In this chapter, we focus on results of quasi-factorization for the Umegaki relative entropy, but weak quasi-factorization has also been proved for the Belavkin--Staszewski (BS) relative entropy \cite{BS82} in \cite{BCP21}. The comparison between that divergence and the Umegaki relative entropy does not by itself compare their marginal differences, so these results cannot simply be substituted into the inequalities above.

In fact, \cite[Section~3]{BCP21} shows that a marginal-difference quasi-factorization for the BS relative entropy cannot hold for all states with only a multiplicative factor $c(\sigma_{AB})$ satisfying $c(\sigma_A\otimes\sigma_B)=1$. For a product reference, expanding the marginal differences makes such a bound equivalent to
\[
 \bs{\rho_{AB}}{\sigma_A\otimes\sigma_B}
 \geq\bs{\rho_A}{\sigma_A}+\bs{\rho_B}{\sigma_B},
\]
namely superadditivity of the BS relative entropy, which fails in general. Indeed, for faithful reference states, superadditivity together with continuity in the first argument, additivity under tensor products and data processing characterizes the Umegaki relative entropy up to a positive constant \cite[Theorem~1]{WGE17}. The BS relative entropy satisfies the latter three properties, and its agreement with Umegaki on commuting states would fix that constant to one, hence it cannot satisfy superadditivity, since otherwise it would coincide with the Umegaki relative entropy, which it does not in general. Thus an additive correction is necessary if the multiplicative factor is to remain one for product states.

\subsection{Conditional relative entropies by recovery}
\label{subsec:weak-qf-recovery}

For recovery maps, we can retain the logarithmic remainder in the proof of Theorem~\ref{thm:qf-recovery-axioms}. This gives weak quasi-factorization directly from the recovery axioms, without requiring a small interaction coefficient. We apply Lieb's three-matrix trace inequality to general recovery maps satisfying our axioms, following the argument of \cite[Theorem~7 and Remark~13]{CLP18b}, and retain the resulting logarithmic correction.

\subsubsection{A logarithmic defect for general recovery maps}

Let $\mcl{R}=(\mcl{R}_A,\mcl{R}_B)$ be the recovery families for the two complement algebras, with fixed-reference maps $\mcl{R}_{A,\sigma_{AB}}$ and $\mcl{R}_{B,\sigma_{AB}}$. Before estimating the interaction term, the proof gives
\begin{equation}\label{eq:weak-qf-recovery-trace}
 \begin{split}
 \ent{\rho_{AB}}{\sigma_{AB}}
 \leq\entXA{\mcl{R}_A}{A}{\rho_{AB}}{\sigma_{AB}}&
 +\entXA{\mcl{R}_B}{B}{\rho_{AB}}{\sigma_{AB}}\\
 &\quad+\log\Tr\left[
 \mcl{R}_{A,\sigma_{AB}}(\rho_{AB})
 \mcl{T}_{\sigma_{AB}}\bigl(\mcl{R}_{B,\sigma_{AB}}(\rho_{AB})\bigr)
 \right].
 \end{split}
\end{equation}
Centering both recovered states at $\sigma_{AB}$ yields the following explicit bound.

\begin{corollary}\label{cor:weak-qf-recovery}
For the maps and coefficient of Theorem~\ref{thm:qf-recovery-axioms},
\begin{equation}\label{eq:weak-qf-recovery}
 \begin{split}
 \ent{\rho_{AB}}{\sigma_{AB}}
 \leq\entXA{\mcl{R}_A}{A}{\rho_{AB}}{\sigma_{AB}}
     & +\entXA{\mcl{R}_B}{B}{\rho_{AB}}{\sigma_{AB}}\\
 &\quad+\log\left(1+\eta_{\mcl{R}}(\sigma_{AB})
 \norm{\rho_A-\sigma_A}_1\norm{\rho_B-\sigma_B}_1\right).
 \end{split}
\end{equation}
In particular, the additive correction can be bounded uniformly over all input states by $\log(1+4\eta_{\mcl{R}}(\sigma_{AB}))$.
\end{corollary}
\begin{proof}
In the proof of Theorem~\ref{thm:qf-recovery-axioms}, retain the logarithm instead of applying $\log u\leq u-1$. The same centering argument bounds its argument by
$1+\eta_{\mcl{R}}(\sigma_{AB})\norm{\rho_A-\sigma_A}_1\norm{\rho_B-\sigma_B}_1$.
Each trace norm is at most two, which proves the uniform bound.
\end{proof}

Corollary~\ref{cor:weak-qf-recovery} can provide the factorization step in a proof of a weak MLSI, following the strategy used for conditional expectations in \cite[Section~4.1]{BCR22}. Suppose a quantum Markov semigroup with invariant state $\sigma_{AB}$ and evolved states $\rho_t$ satisfies the entropy-production estimate
\[
 -\frac{d}{dt}\ent{\rho_t}{\sigma_{AB}}
 \geq\alpha\left(
 \entXA{\mcl{R}_A}{A}{\rho_t}{\sigma_{AB}}
 +\entXA{\mcl{R}_B}{B}{\rho_t}{\sigma_{AB}}\right)
\]
for all initial states and some $\alpha>0$. With $d=\log(1+4\eta_{\mcl{R}}(\sigma_{AB}))$, the corollary gives the weak MLSI
$-\frac{d}{dt}\ent{\rho_t}{\sigma_{AB}}\geq\alpha(\ent{\rho_t}{\sigma_{AB}}-d)$, and hence
\[
 \ent{\rho_t}{\sigma_{AB}}
 \leq e^{-\alpha t}\ent{\rho_0}{\sigma_{AB}}
 +(1-e^{-\alpha t})d.
\]
This controls relaxation down to the entropy scale $d$, even when $\eta_{\mcl{R}}(\sigma_{AB})\geq1/2$ and Theorem~\ref{thm:qf-recovery-axioms} gives no positive multiplicative bound. The entropy-production comparison is an additional assumption on the dynamics that must be verified separately.

\subsubsection{Standard, rotated and averaged Petz maps}

For the standard Petz maps, the estimate
$\eta_{(\mcl{R}_A^P,\mcl{R}_B^P)}(\sigma_{AB})\leq\xi_P(\sigma_{AB})/2$
from Theorem~\ref{thm:qf-petz-explicit} gives
\begin{equation}\label{eq:weak-qf-petz}
 \ent{\rho_{AB}}{\sigma_{AB}}
 \leq\entXA{P}{A}{\rho_{AB}}{\sigma_{AB}}
 +\entXA{P}{B}{\rho_{AB}}{\sigma_{AB}}
 +\log\left(1+2\xi_P(\sigma_{AB})\right).
\end{equation}
This consequence remains valid when $\xi_P\geq1$, where the multiplicative estimate \eqref{eq:qf-petz-explicit} no longer gives a positive lower coefficient.

For averages $\overline{\mcl{R}}_A,\overline{\mcl{R}}_B$ of admissible families with probability measures $\mu,\nu$, set
\[
 \overline\eta(\sigma_{AB})
 =\int\!\!\int\eta_{(\mcl{R}_A^t,\mcl{R}_B^s)}(\sigma_{AB})\,d\mu(t)\,d\nu(s).
\]
Whenever this integral is finite, \eqref{eq:qf-averaged-recovery} gives the uniform defect
\[
 d=\log\left(1+4\overline\eta(\sigma_{AB})\right)
\]
for the averaged maps. This includes the averaged rotated Petz maps with $d\mu(t)=\beta(t)\,dt$ and $d\nu(s)=\beta(s)\,ds$. At a product reference the interaction coefficient vanishes for every admissible family, and the exact product and multipartite results of the previous section apply.

\subsubsection{Finite covers and a tradeoff between the two constants}

The finite-cover result of Proposition~\ref{prop:qf-recovery-finite} also gives a simple uniform weak estimate. If its multiplicative constant is enlarged, if necessary, to $C\geq1$, then
\begin{equation}\label{eq:weak-qf-recovery-cover}
 \ent{\rho}{\sigma}
 \leq\sum_{j=1}^m\entXA{\mcl{R}_j}{A_j}{\rho}{\sigma}
 +\left(1-\frac1C\right)\log\lambda_{\min}(\sigma)^{-1}.
\end{equation}
Indeed, the strong estimate bounds the sum below by $\ent{\rho}{\sigma}/C$, while $\ent{\rho}{\sigma}\leq\log\lambda_{\min}(\sigma)^{-1}$. Subtracting the sum proves \eqref{eq:weak-qf-recovery-cover}. This elementary tradeoff applies to any of the three notions whenever a strong global factorization is available, but the resulting defect can grow with system size.

Finally, the recovery-error estimates of \cite{junge_universal_2018,SBT17}, recalled in Subsection~\ref{subsec:recovery-lower-bounds}, concern measured relative entropy or fidelity. They do not directly give upper bounds on the Umegaki relative entropy to the recovered state, and hence do not supply further quasi-factorization inequalities of the form considered here.


\subsection*{Acknowledgment}
A.C. acknowledges the support of the Deutsche Forschungsgemeinschaft (DFG, German Research Foundation) - Project-ID 470903074 - TRR 352.  This
project was funded within the QuantERA II Programme that has received funding from the EU’s H2020 research and innovation programme under the GA No 101017733.
A.L. acknowledges support from the Italian Ministry of University and Research (MUR), through ``Programma per Giovani Ricercatori Rita Levi Montalcini'', the grant ``Dipartimento di Eccellenza 2023-2027'' of Dipartimento di Matematica, Politecnico di Milano, as well as the National Group of Mathematical Physics (GNFM) of the Italian Institute for High Mathematics (INdAM).
The authors would like to thank the Isaac Newton Institute for Mathematical Sciences, Cambridge, for support and hospitality during the programme Mathematics of Many-Body Entanglement, where work on this chapter was undertaken. This work was supported by EPSRC grant EP/Z000580/1.

\printbibliography

\end{document}